\documentclass[aps,prx,amsmath,twocolumn,amssymb,titlepage]{revtex4-1}

\usepackage{graphicx}
\usepackage{nicefrac}
\usepackage{amsfonts}

\usepackage{amssymb}
\usepackage{amsmath} 
\usepackage{hhline}
\usepackage{subfig}

\usepackage{multirow} 
\usepackage{tabularx} 
\usepackage{array}

\usepackage{units}

\usepackage{color} 

\usepackage{bm}
\usepackage{hyperref}
\usepackage{comment}

\usepackage{xr}
\newcommand{\grafe}[1]{\left\{ #1 \right\}}
\newcommand{\tonde}[1]{\left( #1 \right)}
\newcommand{\quadre}[1]{\left[ #1 \right]}

\makeatletter
\renewcommand\@makecaption[2]{%
  \par
  \vskip\abovecaptionskip
  \begingroup
   \small\rmfamily
    \begingroup
     \samepage
     \flushing
     \let\footnote\@footnotemark@gobble
     \@make@capt@title{#1}{#2}\par
    \endgroup
  \endgroup
  \vskip\belowcaptionskip
}
\makeatother

\begin{document}
\title{Complexity of Energy Barriers in Mean-Field Glassy Systems}
\author{Valentina Ros}
\affiliation{Institut de physique th\'eorique, Universit\'e Paris Saclay, CEA, CNRS, F-91191 Gif-sur-Yvette, France}
\author{Giulio Biroli}
\affiliation{Institut de physique th\'eorique, Universit\'e Paris Saclay, CEA, CNRS, F-91191 Gif-sur-Yvette, France}
\affiliation{Laboratoire de Physique Statistique, Ecole Normale Sup\'erieure,
PSL Research University, 24 rue Lhomond, 75005 Paris, France}
\author{Chiara Cammarota}
\affiliation{King's College London, Department of Mathematics, Strand, London WC2R
2LS, United Kingdom}
% For each author

\begin{abstract}
\noindent
We analyze the energy barriers that allow escapes from a given local minimum in a mean-field model of glasses. 
We perform this study by using the Kac-Rice method and computing the typical number of critical points of the energy function at a given distance from the minimum. We analyze their Hessian in terms of random matrix theory and show that for a certain regime of energies and distances critical points are index-one saddles and are associated to barriers. We find that the lowest barrier, important for activated dynamics at low temperature, is strictly lower than the ``threshold" level above which saddles proliferate. 
We characterize how the quenched complexity of barriers, important for activated process at finite temperature, depends on the energy of the barrier, the energy of the initial minimum, and the distance between them.  
The overall picture gained from this study is expected to hold generically for mean-field models of the glass transition. 
\end{abstract}

\maketitle
\noindent
Many complex systems in physics, biology and computer science are characterized by high-dimensional landscapes full of local minima and saddles of any order. Characterizing the statistical properties of critical points in these energy landscapes is instrumental to explain and predict the static and dynamic behavior of such systems \cite{kurchan6,birolileshouches}.\\
%Metastability is responsible for the slow dynamics of these systems, which at low temperatures remain trapped for long time in the vicinity of some local minimum, occasionally jumping to adjacent minima through the crossing of the energy barriers. These transitions are expected to occur via thermally activated processes, the rate of which depends on the distribution of the heights of the barriers surrounding the local minima of the landscape.\\
Much of the current understanding of this problem comes from the research on the glass transition and spin-glasses, which played a major role in developing methods to study the generic properties of rough high-dimensional landscapes. Several numerical investigations have introduced ways to map out the network of local minima of the potential energy landscape associated to models of glass-formers, and to characterize their properties~\cite{Wales, Heuer, Heuer2, Heuer3, SastriNature, Sastri2}; whereas theoretical works, started in the 80s with the development of spin-glass theory \cite{moore,kurchan,crisom95,CGPConstrainedComplexity}, have obtained the number of critical points and local minima in mean-field models of glasses.  
%This has been found to scale with the number of degrees of freedom (the landscape dimension). 
Recently, a regain of interest on this subject is coming from computer science, and in particular machine learning \cite{someCSreview} where many central questions concern the statistical properties of rough high-dimensional landscapes originating from the study of the multi-dimensional profile of loss functions. Concomitantly, advances in probability theory and mathematical physics are currently allowing to put the theoretical physics methods 
on a firmer basis and to obtain new results \cite{fyodorov,fyodorovnadal,braydean,auffingerbenaouscerny,
subag,MontanariBenArous,touboul,fyodorovmay,SpikedRepKacRice,Ipsen, montanariTAP}.\\
Despite this great amount of progress on enumerating and classifying local minima, the characterisation of the typical {\it energy barriers} between them is still to a large extent an open question. 
In particular, notwithstanding numerical~\cite{Angelani,Broderix,Doyesaddles} and theoretical ~\cite{CGPConstrainedComplexity, CGPThreeReplicas, CGPBarriersThreeReplicas,BarratFranz} attempts in the context of the glass transition, the lack of information about barriers in rough-landscapes remains the main obstacle for the development of a theory of dynamics in glassy systems, and in many other contexts where such landscapes play an important role. \\
In this work we focus on the spherical $p$-spin model~\cite{crisom92} which is 
an archetypical model of rough energy landscapes and of the glass transition \cite{cavagnapedestrian,ReviewBCKM}. Using a method developed in \cite{SpikedRepKacRice}, which builds on the Kac-Rice formula for the computation of stationary points of random functionals~\cite{fyodorov,fyodorovnadal,braydean,auffingerbenaouscerny,
subag,MontanariBenArous,touboul,fyodorovmay}, we work out the full geometrical organization of typical barriers that enable escapes from local minima, obtaining a picture that is expected to generically hold for many glassy systems. 
From the mathematical point of view, this represents a first step towards a full characterization of the Morse complex of random high-dimensional functions. \\ 
%fully characterize the barriers that allow to escape from local minima. By barrier, we mean a critical point, typically a saddle of index one, in the vicinity of a minimum that is temporarily trapping the dynamics. In practice, we compute the typical number of critical points at fixed energy and fixed distance from a given local minimum. We determine the lowest among them and study their Hessian, finding that they play the role of energy barriers associated to the preferred escaping paths at very low temperature. Moreover, we show that beyond a certain distance an large number of them becomes available.  
%Overall, we work out thecfull geometrical organization of barriers obtaining a picture that is expected to hold generically for many glassy systems. \\  
The energy functional of the spherical $p$-spin model reads:
\begin{equation}\label{eq:HampSpin}
 E\quadre{{\bf s}}=-\sum_{\langle i_1,i_2,\dots,i_p\rangle}J_{i_1,i_2,\dots,i_p } s_{i_1} s_{i_2} \dots s_{i_p}, 
\end{equation}
where the sum runs over all the possible $p$-uplets of indexes $i_k$ (going from $1$ to $N$); the configuration ${\bf s}=(s_1, \cdots, s_N)$ lives on an $N$-dimensional hypersphere, \emph{i.e.}, $\sum_{i=1}^{N} s_i^2=N$, and the quenched random couplings $J_{i_1,i_2,\dots,i_p}$ are i.i.d. normally distributed random variables with zero mean and variance $\langle J^2 \rangle =p!/2N^{p-1}$. 
At energy density $\epsilon= \lim_{N \to \infty} E[{\bf s}]/N$ higher than the ground-state, $\epsilon>\epsilon_{\text{gs}}$, the model exhibits a number of stationary points which grows exponentially with the dimension $N$. Their stability changes as a function of $\epsilon$ and can be described in terms of the {\it index}, \emph{i.e.}, the number of downhill directions. 
At high energy the overwhelming majority of critical points are saddles, with an index proportional to $N$. 
At low energy minima are instead exponentially more frequent than saddles~\cite{crisom95, CGPComplexityTAP,auffingerbenaouscerny}.
%, even in comparison with saddles with index of order one. 
The transition between these two regimes is sharp. It occurs at a value of the energy density called \emph{threshold}, $\epsilon_{\text{th}}(p)\equiv-\sqrt{2 (p-1)/p}$, at which typical critical points are characterized by plenty of directions with an almost zero curvature. 
%, since the many downhill directions flatten out before becoming uphill in the low energy range full of minima.   
Low-temperature dynamics of the $p$-spin model starting from high-energy initial conditions is essentially a weak-noise dynamical descent in the energy landscape  (it becomes a gradient descent in the limit of zero temperature). At small temperatures, penetrating below the threshold and reaching the equilibrium energy requires time scales that grow exponentially with $N$~\cite{CuKuPRL,montanarisemerjian,benarousj}. 
On these extremely long time-scales the system decreases its energy by escaping from local minima via index-one saddles, \emph{i.e.}, {\it crossing barriers}. 
This dynamical regime has been studied numerically for some mean-field glassy models in~\cite{CrisantiRitort, CrisantiRitort2,Jorgetrap,MBJ,MBJ2}. Rigorous results have been obtained for the Random Energy Model \cite{benarousrem,cerny,Gayrard}. In order to develop a theory of activated dynamics in this and more complicated settings, it is crucial to understand how barriers are organized in configuration space. Pioneering works addressed this problem for mean-field glass systems like the $p$-spin at the end of the 90s~\cite{CGPConstrainedComplexity, CGPThreeReplicas, CGPBarriersThreeReplicas,BarratFranz,LopatinIoffe}. However the task proved to be so challenging that many central questions remained unanswered. 
For instance, it is still unknown whether the system has
to climb up to the threshold to escape from local minima or can instead sneak through selected paths that involve lower barriers.\\  
Our goal is to address this and similar issues 
%on the number and geometrical disposition of barriers 
by the quenched Kac-Rice formalism we developed in \cite{SpikedRepKacRice}.
Our starting point is the computation of the typical number of saddles surrounding a given minimum. 
This problem was already addressed in 
\cite{CGPConstrainedComplexity} but in a simpler setting.
For convenience, we re-define the state variables setting them on the unit sphere, ${\bm \sigma}= {\bf s}/\sqrt{N}$, and introduce the rescaled energy $h[{\bm \sigma}] \equiv  \sqrt{{2}/{N}} E[\sqrt{N} {\bm \sigma}]$~\cite{fn:EnergyDensity}.
We denote with ${\bf g} \quadre{{\bm \sigma}}$ and $\mathcal{H}\quadre{{\bm \sigma}}$ its gradient vector and Hessian matrix, respectively \cite{fn:TangentPlane}. 
%\begin{figure}[!htbp]
%\includegraphics[scale=0.5]{Figure/Immagine1}
%    \caption{Some sketchy figure for the setup? [this is just placeholder] Maybe the one with equidistant replicas? Maybe one with a tangent plane? Maybe ruba solo spazio: vedere. Magari mettere anche Hessiano? O piano tangente e la direzione che punta verso il minimo?}\label{fig:Immagine1}
% \end{figure}%
% \noindent
  We take a fixed minimum ${\bm \sigma}^0$ drawn at random from the population of minima with energy $\epsilon_0$ ($\epsilon_{\text{gs}} \leq \epsilon_0 \leq \epsilon_{\text{th}}$), and define the number $\mathcal{N}_{{\bm \sigma}^0}(\epsilon, q| \epsilon_0)$ of stationary points with energy $\epsilon$ that are at fixed distance from ${\bm \sigma}^0$, measured by one minus the overlap ${\bm \sigma}^0 \cdot {\bm \sigma}= q$ (high overlap corresponds to small distance). 
\begin{figure}[!ht]
\includegraphics[width=\columnwidth]{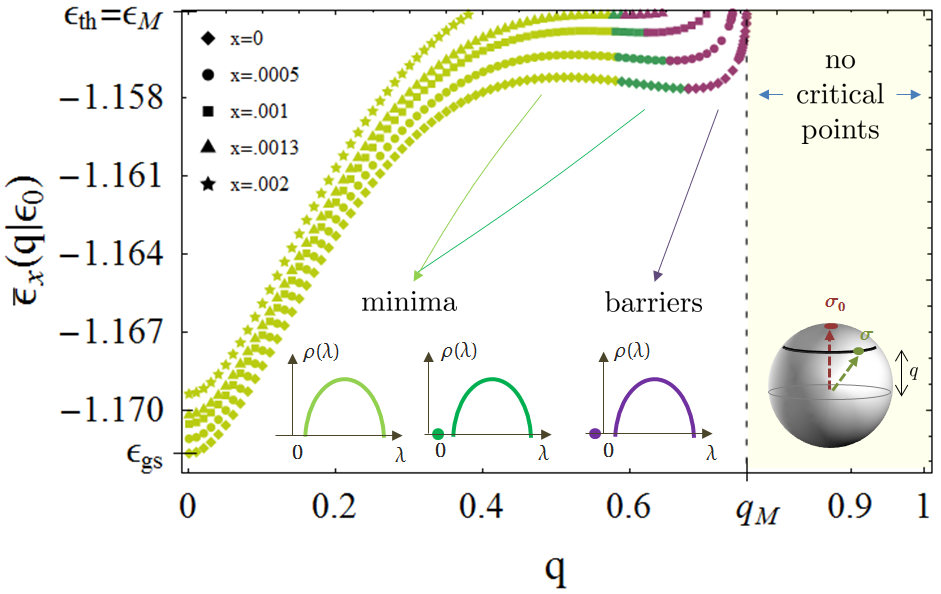}
    \caption{Energy densities $\overline{\epsilon}_x(q|\epsilon_0)$ of the stationary points at overlap $q$ from the fixed minimum and having complexity $\Sigma=x$, for $p=3$ and $\epsilon_0=-1.167$. The green points correspond to minima and the violet points to index-one saddles. The evolution of the density of states of the Hessian is sketched below. Above the threshold energy all points are therefore saddles of index proportional to $N$.}\label{fig:Fig1}
 \end{figure}%
Since saddles above $\epsilon_{\text{th}}$ are not the ones used by activated dynamics, we restrict $\epsilon$ to the same range as $\epsilon_0$.
$\mathcal{N}_{{\bm \sigma}^0}(\epsilon, q| \epsilon_0)$ is a random variable, and we are interested in its {\it typical} value whose logarithm is given by 
%The variable $\mathcal{N}_{{\bm \sigma}^0}(\epsilon, q| \epsilon_0)$ fluctuates -?-at the exponential scale-?- in $N$; information on the landscape in the vicinity of ${\bm \sigma}^0$ can be extracted from its typical value, whose leading-order scaling is governed by 
the {\it quenched} constrained complexity:
\begin{equation}\label{eq:QuenchedConstrained}
 \Sigma(\epsilon, q| \epsilon_0)= \lim_{N \to \infty}\frac{1}{N}\Big\langle \,  \log \mathcal{N}_{{\bm \sigma}^0}(\epsilon, q| \epsilon_0) 
 \Big\rangle_{0},
\end{equation}
where the average is taken over the disorder and the local minima of energy $\epsilon_0$.
Its {\it annealed} counterpart given by $\ln \left\langle \mathcal{N}_{{\bm \sigma}^0}\right\rangle_0$ can be used as an approximation and is accessible to rigorous treatments 
but it coincides with $\Sigma(\epsilon, q| \epsilon_0)$ in a few cases only \cite{auffingerbenaouscerny,
subag} (when the distribution of $\mathcal{N}_{{\bm \sigma}^0}$ concentrates around its average).
The calculation of the quenched complexity follows the method developed recently in \cite{SpikedRepKacRice}. We report below the results and we refer to the Supplemental Material (SM) for the detailed computation and extensions of them. %above $\epsilon_{\text{th}}$. 
%Henceforth, since we are interested in barriers to escape from local minima, we only focus on $\epsilon_{\text{gs}} \leq \epsilon_0 \leq \epsilon_{\text{th}}$ where $\epsilon_{\text{gs}}$ is the typical ground-state energy density of \eqref{eq:HampSpin}.
%We also restrict $\epsilon$ to the same range since saddles above $\epsilon_{\text{th}}$ are not the ones used by activated dynamics (see the SM for the results outside this range).
%is more involved, as it requires to determine the scaling of arbitrarily high moments $ \left\langle \mathcal{N}^n_{{\bm \sigma}^0}\right\rangle_0$, from which the complexity can be obtained via the replica trick. The asymptotic scaling of the moments can be obtained expressing them via a replicated version of the Kac-Rice formula~\eqref{eq:FirstMomentKacRice}, see the Supplemental Material (SM).
% through the analytic continuation:
%\begin{equation}\label{eq:ReplicaTrick}
% \Sigma(\epsilon, q |\epsilon_0)= \lim_{N \to \infty}\lim_{n \to 0}\frac{\left\langle \mathcal{N}^n_{{\bm \sigma}^0}(\epsilon, q| \epsilon_0)\right\rangle_0-1}{Nn}. 
%\end{equation}
%In the Supplemental Material (SM), we express the higher moments making use of a replicated version of the Kac-Rice formula \eqref{eq:FirstMomentKacRice}, extract their leading order behavior in $Nn$ and perform the analytic continuation to get the complexity. 
The quenched complexity reads:
\begin{equation}\label{eq:FinalComplexityPostSaddle}
 \Sigma(\epsilon,q| \epsilon_0) =  \frac{1}{2}              \log\tonde{\frac{p}{2} \tonde{\tilde{z}-\epsilon}^2} + \frac{p \tonde{\epsilon^2+ \epsilon \tilde{z}}}{2(p-1)} + \frac{Q}{2}
 \end{equation}
 where
 \begin{equation*}       
           Q= \log \tonde{\frac{1-q^2}{1-q^{2p-2}}}- 2\tonde{\epsilon_0^2 U_0(q)+ \epsilon_0 \epsilon U(q)+ \epsilon^2 U_1(q)},
\end{equation*}
with $\tilde{z}=\sqrt{\epsilon^2- \epsilon_{\text{th}}^2}$ and
\begin{equation*}
 \begin{split}
  U_0(q)&=\frac{q^{2 p} \left(-q^{2 p}+p \left(q^2-q^4\right)+q^4\right)}{q^{4 p}-\left((p-1)^2 (1+q^4)-2 (p-2) p q^2\right) q^{2 p}+q^4},\\
  U(q)&=\frac{2 q^{3 p} \left(p \left(q^2-1\right)+1\right)-2 q^{p+4}}{q^{4 p}-\left((p-1)^2 (1+q^4)-2 (p-2) p q^2\right) q^{2 p}+q^4},\\
  U_1(q)&=\frac{q^4-q^{2 p} \left(p \left((p-1) q^4+(3-2 p) q^2+p-2\right)+1\right)}{q^{4 p}-\left((p-1)^2 (1+q^4)-2 (p-2) p q^2\right) q^{2 p}+q^4}.
 \end{split}
\end{equation*}
%coefficients $U_i(q)$ given in the SM.
For $q=0$, one finds $U_0=U=0$, $U_1=1$, and $Q=-2\epsilon^2$ and we recover the expression of the unconstrained complexity 
$\Sigma(\epsilon)$ \cite{crisom95}, which counts the typical number of stationary points irrespectively of their location in the space of configurations.
%This unconstrained limit is denoted by $\Sigma(\epsilon)$; it is independent of $\epsilon_0$ and corresponds to $U_0=U=0$, $U_1=1$, and $Q=-\epsilon^2$.
%It becomes independent of $\epsilon_0$, \emph{i.e.} $U_0=U=0$ and $Q=-\epsilon^2$ and will be simply denoted by $\Sigma(\epsilon)$.  
The expression \eqref{eq:FinalComplexityPostSaddle} for the constrained quenched complexity turns out to be equal to the one of the annealed complexity.
\begin{figure}[!htbp]
\includegraphics[width=\columnwidth]{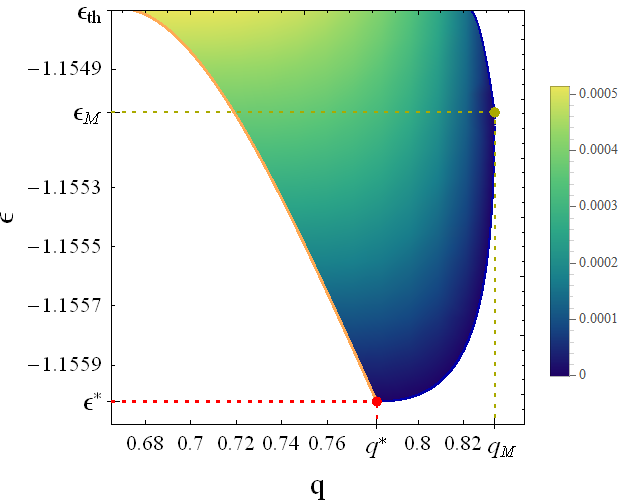}
    \caption{Barriers complexity as a function of $q$ and $\epsilon$ for $p=3$ and $\epsilon_0=-1.158$. Note that in this case $\epsilon_M<\epsilon_{\text{th}}$. %Complexity of the saddles of index one for different values of their overlap $q$ with the fixed minimum. For the smaller $q$, minima and saddle coexist: the dots denote the energies at which the isolated eigenvalue of the Hessian becomes equal to zero, indicating that stationary points with smaller $\epsilon$ are minima (their complexity is not plotted). For this $\epsilon_0$, $q_{\text{M}}(\epsilon_0)=0.7587$ (the violet curve corresponding to a larger overlap is indeed entirely below zero). The saddles with lowest energy are found at $\overline{q}(\epsilon_0)=0.679$, with $\epsilon^*(\epsilon_0)= -1.15763$.  
    }\label{fig:Fig2}
 \end{figure}%
\noindent
This is quite surprising since the presence of the constraint was expected to lead to non-trivial correlations
between critical points, and hence to a difference between quenched and annealed averages. It is a fortunate
coincidence though, since it simplifies considerably the analysis of the Hessian, it opens the way to a rigorous proof of (\ref{eq:FinalComplexityPostSaddle}), along the lines of~\cite{subag} and, moreover, 
%We find that for any $q$ the expression \eqref{eq:FinalComplexityPostSaddle} reproduces exactly the annealed complexity given by  
%$\ln \left\langle \mathcal{N}_{{\bm \sigma}^0}\right\rangle_0$, implying that in the large $N$ limit the distribution of $\mathcal{N}_{{\bm \sigma}^0}$ concentrates around its average. 
%This is a quite unexpected result  in presence of the constraint on the overlap. 
%It opens the way to a rigorous proof of (\ref{eq:FinalComplexityPostSaddle}), along the lines of~\cite{subag}. 
it justifies \emph{a posteriori} the annealed approximation of \cite{CGPConstrainedComplexity}.\\ 
%Note that with respect to \cite{CGPConstrainedComplexity}, our analysis provides very detailed information on the Hessian of the critical points and hence allows us to discuss the geometrical organization of barriers, as we show below.  
We now focus on the properties of the Hessian of the critical points.
Depending on the values of $q, \epsilon$ and $\epsilon_0$, we find that the points
counted by \eqref{eq:FinalComplexityPostSaddle} are either minima or saddles with one unstable direction 
as long as $\epsilon<\epsilon_{\text{th}}$. 
More precisely the matrices $\mathcal{H}[{\bm \sigma}]$ are
%More precisely, the matrix $\mathcal{H}[{\bm \sigma}]$ at a typical stationary point ${\bm \sigma}$ with energy $\epsilon$ and at fixed overlap $q$ from the minimum ${\bm \sigma}^0$ is 
distributed as $(N-1)\times (N-1)$ GOE matrices with variance $\sigma^2=p(p-1)$, 
%~\cite{fn:ShiftGOE} 
perturbed by a diagonal matrix with entries equal to $\sqrt{2 N}p \epsilon$ and 
by a rank-one matrix, that depends on $q, \epsilon$ and $\epsilon_0$. The corresponding bulk eigenvalues density is therefore a shifted semicircle law with a positive support whose lower edge touches zero for $\epsilon\rightarrow\epsilon_{\text{th}}^-$. 
%Its lower edge touches zero only at the threshold energy.
%At the threshold energy the lower edge of the semicircle touches zero. 
Interestingly, the rank-one perturbation can push an isolated eigenvalue out from the semicircle  %and can even become negative for $\epsilon<\epsilon_{\text{th}}$ 
for certain values of parameters (see Fig.~\ref{fig:Fig1}) \cite{jones,bbp}. %When it exists, the eigenvalue reads:
When this happens, the expression of the isolated eigenvalue reads
\begin{equation}
 \lambda_0(q, \epsilon, \epsilon_0)= \frac{\mu \tonde{1\hspace{-.05 cm}-\hspace{-.05 cm}\frac{\Delta^2}{2 \sigma^2}}+ \frac{\Delta^2}{\sigma} \sqrt{\hspace{-.05 cm}\frac{\mu^2}{4 \sigma^2}\hspace{-.05 cm}-\hspace{-.05 cm} \tonde{1\hspace{-.05 cm}-\hspace{-.05 cm}\frac{\Delta^2}{\sigma^2}}}}{(1-\Delta^2/\sigma^2)}- \sqrt{2} p \epsilon \ ,
\end{equation}
where the expressions of $ \mu(q, \epsilon, \epsilon_0)$ and $\Delta^2(q)$ are given in the SM.
%\begin{equation*}\label{eq:Mu}
% \begin{split}
% \mu(q, \epsilon, \epsilon_0) \equiv\frac{\sqrt{2} (p-1) p \left(1-q^2\right) \left(a_0(q) \epsilon_0-a_1(q) \epsilon \right)}{a_2(q)}
% \end{split}
%\end{equation*} 
%with 
%\begin{equation*}
%\begin{split}
% a_0&= q^4+ q^{2 p}\left(1-p +(p-2)q^2\right),\\
% a_1&=q^{3 p}+ q^{p+2}\left(p-2-(p-1) q^2\right),\\
% a_2&=q^{6-p}+q^{3 p+2}- q^{p+2}\left((p-1)^2 (q^4+1)-2 (p-2) p q^2\right),
%   \end{split}
%\end{equation*}
%and 
%\begin{equation*}
% \Delta^2(q)=p(p-1)\quadre{1-\frac{(p-1)(1-q^2) q^{2p-4}}{1-q^{2p-2}}}.
%\end{equation*}
Its corresponding eigenvector has a finite projection on the direction that points toward ${\bm \sigma}^0$. 
Points for which $\lambda_0<0$ are saddles having one unstable direction connecting ${\bm \sigma}$ with ${\bm \sigma}^0$. Hence, they correspond to possible ``mountain-passes" to escape from ${\bm \sigma}^0$.    
 Henceforth, we will call them barriers. \\%: they are at the top of the barriers that are crossed dynamically to escape from the minimum ${\bm \sigma}^{0}$. \\
%In order to present our results, we first consider $\epsilon_0$ fixed. 
\noindent
\begin{figure}[!htbp]
\includegraphics[width=\columnwidth]{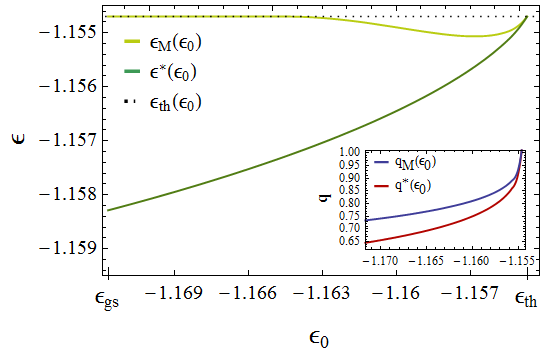}
    \caption{Energy $\epsilon_{\text{M}}$ of the closest saddles and energy $\epsilon^*$ of the minimal-energy ones, as a function of $\epsilon_0 \in \quadre{\epsilon_{\text{gs}}, \epsilon_{\text{th}}}$.  For $-1.16421 \leq\epsilon_0 \leq \epsilon_{\text{th}}$, the closest saddles have energy that is below the threshold (black dotted line). {\it Inset. }Latitudes $q_{\text{M}}(\epsilon_0)$ of the closest saddles and latitude $q^*(\epsilon_0)$ of the ones of minimal energy.
}\label{fig:Fig3}
\end{figure}
\noindent
In Fig.~\ref{fig:Fig1} we show iso-complexity energy curves $\overline{\epsilon}_x(q|\epsilon_0)$ defined by $\Sigma(\overline{\epsilon}_x, q|\epsilon_0)=x$ for fixed $\epsilon_0$: $\overline{\epsilon}_x(q|\epsilon_0)$
is the energy of typical stationary points with overlap with the reference minimum equal to $q$ and 
complexity equal to $x$.
%In Fig.~\ref{fig:Fig1} we show the energy $\overline{\epsilon}_x(q|\epsilon_0)$ of the typical stationary points with complexity equal to $x$ and at overlap $q$ with a minimum ${\bm \sigma}^{0}$ for a given fixed $\epsilon_0$.
%In Fig.~\ref{fig:Fig1} we show iso-complexity energy curves $\overline{\epsilon}_x(q|\epsilon_0)$ defined by $\Sigma(\overline{\epsilon}_x, q|\epsilon_0)=x$ for fixed $\epsilon_0$. 
%$\overline{\epsilon}_x(q|\epsilon_0)$ is the energy of the typical stationary points at overlap $q$ with ${\bm \sigma}^{0}$ that have complexity equal to $x$. 
%The $x=0$ curve is the lowest one in energy, since as 
%The picture emerging from Fig.~\ref{fig:Fig1} is the following: 
The study of iso-complexity curves shows that at high $q$ the energy landscape is convex. Critical points other than ${\bm \sigma}^{0}$ only appear beyond a minimal distance from ${\bm \sigma}^{0}$ (\emph{i.e.}, $q<q_{\text{M}}$), and 
usually at the threshold energy, or at an energy $\epsilon_M$ slightly below (see later). When they appear, they are barriers. 
Increasing the distance, the isolated eigenvalue grows until it becomes positive, and critical points become minima. On the iso-complexity curves 
this happens when $\overline{\epsilon}_x(q|\epsilon_0)$ reaches a local minimum (change from purple to green in Fig.~\ref{fig:Fig1}). 
We do not have any intuitive explanation of this intriguing coincidence, but we recall that the non-monotonic 
dependence of $\overline{\epsilon}_x(q|\epsilon_0)$ was already noted in~\cite{CGPComplexityTAP} for $x=0$.
At even larger distances, the isolated eigenvalue enters into the semi-circle. Eventually, at $q=0$, we recover the unconstrained complexity result. 
%This is expected on general grounds since the portion of configuration space that corresponds to $q=0$ is exponentially larger (in $N$) than the one at any finite $q$; therefore, the exponential majority of minima of \eqref{eq:HampSpin} lies in this region. 
Among the different curves in Fig.~\ref{fig:Fig1}, the lowest one corresponding to $x=0$ is of particular interest since it gives the typical energy of the deepest stationary points found at overlap $q$ with ${\bm \sigma}^0$. Its local minimum at high overlap, which we denote by ($q^*, \epsilon^*$), represents the lowest energy barrier that can be used to escape from ${\bm \sigma}^0$.\\
 From this analysis at fixed $\epsilon_0$, two relevant information on the landscape can be deduced: (i) there exists a minimal energy barrier that the system has to cross dynamically to exit from the minimum ${\bm \sigma}^0$. 
 This {\it optimal barrier}, which is generically lower 
than $\epsilon_{\text{th}}$, is the one relevant for activated dynamics at very low temperature \cite{fn:optimal}; (ii) there is an exponential number of higher energy barriers. These are relevant for slow dynamics at finite temperature, where higher but more numerous barriers are favored \cite{kurchan6}.  
\noindent
\begin{figure}[!htbp]
\includegraphics[width=\columnwidth]{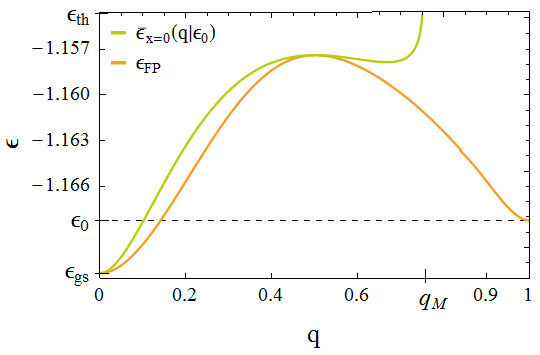}
    \caption{Comparison between the energy density $\overline{\epsilon}_{x=0}$ of the critical points with $\Sigma=0$, and the zero-temperature FP potential $\epsilon_{\text{FP}}$, for $p=3$ and $\epsilon_0=-1.1682.$ 
}\label{fig:Fig4}
\end{figure}
\noindent
Their organization and their complexity is shown in Fig.~\ref{fig:Fig2}.
At high $q$, barriers have high energies and low complexity. Note that at the considered $\epsilon_0$, we have $\epsilon_M<\epsilon_{\text{th}}$. The spectrum of possible 
energies is maximal at the $q$ corresponding to optimal barriers. It then shrinks to zero at low $q$, in correspondence to the highest and most numerous barriers. \\
This scenario depends on $\epsilon_0$ in the following way. 
The energy of the optimal barriers, $\epsilon^*(\epsilon_0)$, remains always below the threshold, and converges to it only when $\epsilon_0\rightarrow\epsilon_{\text{th}}^-$, see Fig. 3. 
A comparison with~\cite{CGPComplexityTAP,auffingerbenaouscerny} shows that the energy of the optimal barriers, despite being below the threshold, is nevertheless much higher that the energy of generic index-one saddles with same complexity as the minima of energy $\epsilon_0$ (see SM for details). This fact points toward a complex geometrical organization of the critical points in phase space. 
As shown in Fig.~\ref{fig:Fig3}, the energy of the closest barriers, $\epsilon_{\text{M}}$, also tends to  $\epsilon_{\text{th}}$ when $\epsilon_0\rightarrow\epsilon_{\text{th}}^-$ however it shows a non-monotonic dependence on $\epsilon_0$~\cite{fn:eMeth}. 
In summary, the spectrum of available barriers is larger for smaller  $\epsilon_0$ and shrinks 
when approaching the threshold, where marginal stationary points are expected to be immediately surrounded by other marginal stationary points \cite{CGPConstrainedComplexity}. This is confirmed in the inset of Fig.~\ref{fig:Fig3}, where 
the overlap of the closest barriers, $q_{\text{M}}(\epsilon_0)$, as well as the overlap of the optimal barriers $q^*(\epsilon_0)$, are shown to approach one 
when $\epsilon_0$ reaches $\epsilon_{\text{th}}$.\\
The curve $\overline{\epsilon}_{x=0}(q|\epsilon_0)$, which measures the deepest stationary points at overlap $q$ from ${\bm \sigma}^0$, shares similarities with the so-called Franz-Parisi (FP)
potential \cite{FP}. Since we are focusing on the energy landscape, and not the free-energy one, the suitable Franz-Parisi potential to compare with is the one at zero temperature: the minimal energy of configurations at overlap $q$ from ${\bm \sigma}^0$. We compute it in the SM using standard replica techniques \cite{FP} and we compare it to $\overline{\epsilon}_{x=0}(q|\epsilon_0)$ in Fig. \ref{fig:Fig4}.  
Since critical points are a subset of all configurations at fixed $q$, the FP potential must be generically lower than or equal to $\overline{\epsilon}_{x=0}(q|\epsilon_0)$. 
%(so the deepest critical point is generically higher in energy than the deepest configuration). The two curves coincide for $q=0$, since in both cases one is focusing on the ground states. 
As shown in Fig. \ref{fig:Fig4}, we find (i) that the two functions are equal only at $q=0$ and at the local maximum of the FP potential, for which the associated critical points are actually {\it minima} (see Fig.\ref{fig:Fig1}), and (ii) that the FP potential is well below $\overline{\epsilon}_{x=0}(q|\epsilon_0)$ for $q=q^{*}$. These results 
show that the FP potential is not directly related to the barriers to escape from ${\bm \sigma}^0$ \cite{fn:top}.
%The coincidence at the local maximum instead comes as a surprise: it means that configurations with overlap corresponding at the local maximum of the FP potential are critical points, actually {\it minima} (see Fig.\ref{fig:Fig1}). This makes clear that the FP potential, in particular its maximum, is not directly related to barriers to escape from ${\bm \sigma}^0$ \cite{fn:top}. 
In order to detect them, generalized three-replica potentials were introduced and studied dynamically \cite{CGPThreeReplicas,CGPBarriersThreeReplicas,BarratFranz}. On the basis of our results, we expect that even those constructions are not able to capture optimal barriers (as suggested comparing their typical overlaps \cite{fn:overlaps}). The physical reason is that within the three-replica potential formalism optimal barriers are atypical and in order to probe them one has to combine it with large deviations techniques as shown in \cite{FranzRocchi}.\\
In conclusion, we have worked out the complex organization of barriers to escape from a given minimum in the $p$-spin spherical model, obtaining a scenario that is expected to hold generically for mean-field disordered systems displaying a glass transition.
%We have identified optimal barriers, relevant at low temperature, and compute the complexity of barriers,
%which plays an important role for activated dynamics at finite temperature. 
%In fact, in this case, higher but more numerous 
%barriers are expected to dominate the escape rate \cite{fn:bar}. 
Our results suggest several other important directions to investigate further. First, it would be important to generalize our computation to locate all the barriers connected 
to the reference minimum, not only when they are typical (as analyzed here) but also when they are subdominant (rare) in comparison with other critical points.
Second, it is interesting to know the properties of the minima (other than the original one) to which the barriers we have identified are connected to. This would give additional information on activated dynamical paths, that are formed by sequence of jumps through optimal barriers and subsequent minima. 
For instance we have found that the optimal barrier to escape a given minimum is placed at an energy that is smaller than the threshold energy. The next important issue to address is finding out whether the lowest optimal barrier encountered through activated paths leading to thermal relaxation (\emph{i.e.}, connecting minima at zero overlap) is also lower than the threshold.  
In this context one important question is whether an effective description
in terms of trap-like dynamics emerges at long times, as discussed for real systems \cite{Heuer2} and found for the random energy model \cite{MBJ,MBJ2,benarousrem,cerny,Gayrard}. 
The extension of our work to finite temperature, that necessitates to consider the free-energy landscape, is another interesting direction. \\
Working out the dynamical theory of activated processes in mean-field glassy systems is arguably 
one of the most important and challenging problem in glass physics. 
The generalization of the instantonic solutions of the Martin-Siggia-Rose field theory found in ~\cite{IoffeSherrington, LopatinIoffe, BiroliKurchan} combined with the knowledge gained in this work on the structure and the organization of barriers in configuration space provide a promising starting point for this enterprise.  

\begin{acknowledgments}
We thank G. Ben Arous, A. Cavagna, S. Franz and J. Rocchi for discussions. 
We are particularly grateful to 
S. Franz and J. Rocchi for stimulating inputs and for sharing with us their unpublished results \cite{FranzRocchi}.  
This work was partially funded by the Simons Foundation collaboration Cracking the Glass Problem (No. 454935 to G. Biroli).
\end{acknowledgments}

  \section*{Supplemental Material}
  
\noindent
This Supplemental Material contains a detailed presentation of the analysis whose results are presented in the main text.\\
  In the first sections of this Supplemental Material (from A to D) we present the derivation of the quenched constrained complexity. The derivation follows closely the one presented in Ref.~\cite{SpikedRepKacRice} for a similar setup, and we refer to that work for results that extend straightforwardly to this case. In Sec.~\ref{app:SaddlePoint} we show that, at the saddle point, the quenched complexity reproduces the annealed result presented in the main text. In Sec.~\ref{app:IsolatedEigenvalue} we derive the stability of the typical stationary points counted by the complexity, through the analysis of the statistical properties of their Hessian. In Sec.~\ref{app:AdditionalResults} we report some additional results on the complexity. Sec.~\ref{app:FranzParisi} contains the calculation of the zero-temperature Franz-Parisi potential.
  
  \subsection{Replicated Kac-Rice formula}\label{app:Replicated}
\noindent
This complexity in Eq. (2) of the main text is {\it quenched} since the disorder average is performed over an intensive quantity (the logarithm of the number $\mathcal{N}_{{\bm \sigma}^0}$), rather that over the number itself. The {\it annealed} version of the complexity is obtained exchanging the disordered average with the logarithm, and can be computed expressing the first moment of $\mathcal{N}_{{\bm \sigma}^0}$ by means of the the Kac-Rice formula. This reads (see also Ref. \cite{CGPConstrainedComplexity}):
 \begin{equation}\label{eq:FirstMomentKacRice}
 \left\langle \mathcal{N}_{{\bm \sigma}^0}\right\rangle_0= \int d{\bm \sigma} \, \delta\hspace{-0.05 cm}\tonde{{\bm \sigma} \hspace{-0.05 cm}\cdot\hspace{-0.05 cm} {\bm \sigma}^0\hspace{-0.05 cm}-\hspace{-0.05 cm} q\hspace{-0.01 cm}} \langle  \left| \text{det} \mathcal{H}[{\bm \sigma}]\right|\rangle_0\, p_{{\bm \sigma}|{\bm \sigma}^0}({\bf 0}, \epsilon),
 \end{equation}
 where the integration is over configurations ${\bm \sigma}$ on the unit sphere, and $ p_{{\bm \sigma}|{\bm \sigma}^0}({\bf 0}, \epsilon)$ is the joint density function of the gradient and field $({\bf g} [{\bm \sigma}], h[{\bm \sigma}])$ evaluated at $({\bf 0}, \sqrt{2 N} \epsilon)$, conditioned to ${\bm \sigma}^0$. The first moment of $ \mathcal{N}_{{\bm \sigma}^0}$ is thus formed by three terms: the joint distribution gives the probability that ${\bm \sigma}$ is a stationary point and accounts for its correlations with ${\bm \sigma}^0$, the expectation value of the determinant counts the multiplicity of stationary points in a given level set of the landscape, and the integration over the volume at fixed overlap $q$ accounts for the phase space available to them. Three analogous contributions appear also in the calculation of the quenched complexity. To perform the averages of the logarithm, we exploit the replica trick:
\begin{equation}
 \Sigma(\epsilon, q| \epsilon_0)= \lim_{N \to \infty}\lim_{n \to 0}\frac{M_n(\epsilon, q| \epsilon_0)-1}{Nn} , 
\end{equation}
where 
\begin{equation}
M_n(\epsilon, q| \epsilon_0)\equiv
 \Big\langle \mathcal{N}^n_{{\bm \sigma}^0}(\epsilon, q| \epsilon_0) \Big|  \grafe{ \begin{subarray}{l}
 {\bf g}[{\bm \sigma}^0]=0,\\
  h[{\bm \sigma}^0]=\sqrt{2 N} \epsilon_0 
  \end{subarray}} \Big\rangle
\end{equation}
is the expression for the higher moments of $\mathcal{N}_{{\bm \sigma}^0}$, which can be obtained by replicating the Kac-Rice formula for the first moment. This involves introducing $n$ configurations ${\bm \sigma}^a$, $a=1, \cdots, n$ (which we henceforth refer to as \emph{replicas}), all at fixed overlap $q$ with ${\bm \sigma}^0$. For all the $n+1$ points labeled \cite{footnote1} by $\alpha=0,1,\cdots,n$ we define the gradient vectors ${\bf g}^\alpha \equiv {\bf g}[{\bm \sigma}^\alpha]$, the Hessian $\mathcal{H}^\alpha\equiv \mathcal{H}[{\bm \sigma}^\alpha]$, and the value of the rescaled energy functional $h^\alpha\equiv h[{\bm \sigma}^\alpha]$ defined in the main text. We denote with $\vec{{\bf g}}=({\bf g}^1, \cdots, {\bf g}^n)$ the $(N-1)n$-dimensional vector collecting the gradients of the $n$ replicas, and with $\vec{ h}=(h^1, \cdots, h^n)$ the collection of the $n$ functionals $h^a$. We let $p_{\vec{{\bm \sigma}}|{\bm \sigma}^0}$ be the joint density function of the gradients $\vec{\bf g}$ and fields $\vec{h}$, induced by the distribution of the couplings  and conditioned to ${\bf g}^0={\bf 0}$ and $h^0= \sqrt{2 N} \epsilon_0$. With this notation, the replicated version of the Kac-Rice formula reads:
 \begin{equation}\label{eq:Full}
 M_n= \int \prod_{a=1}^n d{\bm \sigma}^a \, \delta \tonde{{\bm \sigma}^a \cdot {\bm \sigma}^0- q} \mathcal{E}_{\vec{{\bm \sigma}}|{\bm \sigma}^0}(\epsilon)\, p_{\vec{{\bm \sigma}}|{\bm \sigma}^0}({\bf 0}, \epsilon),
 \end{equation}
 where the integration is over configurations ${\bm \sigma}^a$ constrained to be in the unit sphere, at overlap $q$ with the fixed minimum ${\bm \sigma}^0$.
 In \eqref{eq:Full}, $ p_{\vec{{\bm \sigma}}|{\bm \sigma}^0}({\bf 0}, \epsilon)$ is a shorthand notation for the joint density evaluated at $\vec{{\bf g}}=0$ and $h^a= \sqrt{2 N} \epsilon$ for any $a=1, \cdots, n$, while
 \begin{equation}\label{eq:Expectation}
 \mathcal{E}_{\vec{{\bm \sigma}}|{\bm \sigma}^0}(\epsilon)=  \Big\langle\tonde{\prod_{a=1}^n \left| \text{det}\; \mathcal{H}^a\right|}  \Big|
 \grafe{
 \begin{subarray}{l}
 h^a = \sqrt{2 N}\epsilon, h^0=\sqrt{2 N} \epsilon_0\\ 
 {\bf g}^a={\bf 0} \; \forall  a=0, ...,n \end{subarray}} 
 \Big\rangle
 \end{equation}
 denotes the expectation value of the product of the determinants of the Hessians of all replicas, conditioned on each ${\bm \sigma}^a$ being a stationary point with rescaled energy $\sqrt{2 N} \epsilon$ and overlap $q$ with the stationary point ${\bm \sigma}^0$. \\
To extract the leading order in $N$ of \eqref{eq:Full}, we need to characterize the joint distribution of the energy, gradient and Hessian fields at the points ${\bm \sigma}^a$, conditioned to the presence of ${\bm \sigma}^0$. This involves choosing a set of $n+1$ orthonormal bases $\mathcal{B}[{\bm \sigma}^\alpha]=\grafe{{\bf e}^\alpha_1, \cdots, {\bf e}^\alpha_{N-1}}$ in the tangent planes at each ${\bm \sigma}^\alpha$, and computing the averages and covariances of all the fields components with respect to these bases. As it follows from the isotropy of the covariance field of the $p$-spin Hamiltonian, the resulting correlations depend only on the scalar products ${\bf e}^\alpha_i \cdot {\bf e}^\beta_j$ and ${\bf e}^\alpha_i \cdot {\bm \sigma}^\beta$, see the following section for the explicit expressions. If the bases $\mathcal{B}[{\bm \sigma}^\alpha]$ are chosen suitably, the joint distribution of all fields components can be parametrized only in terms of the mutual overlaps $q_{\alpha \beta}= {\bm \sigma}^\alpha \cdot {\bm \sigma}^\beta$ between all configurations (included the fixed overlap $q_{0 a}=q$ with ${\bm \sigma}^0$). This allows to re-write \eqref{eq:Full} as:
 \begin{equation}\label{eq:Full3}
M_n=\hspace{-0.1 cm}  \int \hspace{-0.15 cm}\prod_{a < b=1}^n \hspace{-0.1 cm} dq_{ab} \,e^{N S_n(\epsilon, \hat{Q}| \epsilon_0)+ o(Nn)} ,
\end{equation}
where the leading-order term at the exponent depends on the ${\bm \sigma}^a$ only through the $(n+1) \times (n+1)$ overlap matrix $\hat{Q}$ with components:
\begin{equation}
 Q_{\alpha \beta}=\delta_{\alpha \beta}+(1-\delta_{\alpha \beta}) \quadre{q_{\alpha \beta} +(\delta_{\alpha 0}+ \delta_{\beta 0})(q-q_{\alpha \beta})}.
\end{equation}
The quenched complexity is determined by the linear term in $n$ of $S_n(\epsilon, \hat{Q}| \epsilon_0)$. We thus set:
\begin{equation}\label{eq:LinearExpansion}
 S_n(\epsilon, \hat{Q}| \epsilon_0)= n \Sigma(\epsilon, \hat{Q}| \epsilon_0)+ O(n^2),
\end{equation}
and derive in Sec.~\ref{app:FullDerivation} the explicit expression of $\Sigma(\epsilon, \hat{Q}| \epsilon_0)$. This is obtained within the ansatz $q_{ab} \equiv q_1$, which corresponds to assuming a 1RSB structure of the landscape in the vicinity of the fixed minimum. The calculation is concluded by performing the integral over $q_1$ with the saddle point method, see Sec.~\ref{app:SaddlePoint}.
   \subsection{Covariances of fields and choice of basis vectors}\label{app:Covariances}
 \noindent
  To evaluate explicitly Eq.~\eqref{eq:Full}, we need to characterize of the joint distribution of the fields $\mathcal{H}^a$, ${\bf g}^a$ and $h^a$. \\
 We remind that ${\bf g}^a$ and $\mathcal{H}^a$ denote the \emph{Riemannian} gradient and Hessian fields, which account for the spherical constraint and which lie in the tangent plane to the sphere at each ${\bm \sigma}^a$.
   For simplicity, for each $\alpha=0, 1, \cdots, n$ we introduce also the gradients ${\bm \nabla} h^\alpha \equiv {\bm \nabla} h[{ {\bm \sigma}^\alpha}]$ and Hessian ${\bm \nabla}^2 h^\alpha \equiv {\bm \nabla}^2 h[{\bm \sigma}^\alpha]$ of the rescaled energy functional \emph{extended} to the whole $N$-dimensional space, and determine the covariances between their components along arbitrary directions in this space, given by some $N$-dimensional unit vectors ${\bf v}_i$. 
 The correlations of the components ${\bf g}^\alpha$ and $\mathcal{H}^\alpha$ of the Riemannian gradients and Hessians are easily determined choosing ${\bf v}_i \to {\bf e}_\beta^\alpha$ to be vectors on the tangent plane at the various ${\bm \sigma}^\alpha$; 
 indeed, ${\bf g}^\alpha$ is an $(N-1)$-dimensional vector with components ${g}^\alpha_\beta= {\bm \nabla} h^\alpha \cdot {\bf e}_\beta^\alpha$, obtained from ${\bm \nabla} h^\alpha$ by a projection onto the tangent plane. 
 Similarly, the Riemannian Hessian $\mathcal{H}^\alpha$ is an $(N-1) \times (N-1)$ matrix whose components are related to the ones of ${\bm \nabla}^2 h^\alpha$ by:  
\begin{equation}\label{eq:relHessians}
 \mathcal{H}^\alpha_{\beta \gamma}= {\bf e}_\beta^\alpha \cdot \tonde{{\bm \nabla}^2 h^\alpha-\tonde{{\bm \nabla} h^\alpha \cdot {\bm \sigma}^\alpha}  \hat{1}} \cdot {\bf e}_\gamma^\alpha.
 \end{equation} 
 For arbitrary ${\bf v}_i$ it holds
 \begin{equation}\label{eq:AvGrad}
 \begin{split}
 & \Big\langle \tonde{{\bm \nabla} h^\alpha \cdot {\bf v}_1} h^\beta  \Big\rangle= p ({\bm \sigma}^\alpha \cdot {\bm \sigma}^\beta)^{p-1} \tonde{{\bf v}_1 \cdot {\bm \sigma}^\beta},\\
  & \Big\langle \hspace{-0.1cm} \tonde{{\bf v}_1 \hspace{-0.1cm}\cdot \hspace{-0.1cm} {\bm \nabla}^2 h^\alpha \hspace{-0.1cm} \cdot \hspace{-0.05cm} {\bf v}_2}  h^\beta \Big\rangle=p(p\hspace{-0.05cm}-\hspace{-0.05cm}1)({\bm \sigma}^\alpha \hspace{-0.05cm}\cdot\hspace{-0.05cm} {\bm \sigma}^\beta)^{p-2}({\bf v}_1 \hspace{-0.05cm}\cdot\hspace{-0.05cm} {\bm \sigma}^\beta) ({\bf v}_2 \hspace{-0.05cm}\cdot\hspace{-0.05cm} {\bm \sigma}^\beta),
  \end{split}
 \end{equation}
 while the covariances between the gradient components read:
 \begin{equation}\label{eq:CovGradComp}
 \begin{split}
  & \Big\langle \tonde{{\bm \nabla} h^\alpha \cdot {\bf v}_1} \tonde{{\bm \nabla} h^\beta \cdot {\bf v}_2} \Big\rangle =p ({\bm \sigma}^\alpha \cdot {\bm \sigma}^\beta)^{p-1} \tonde{{\bf v}_1 \cdot {\bf v}_2}+\\
  &p(p-1)({\bm \sigma}^\alpha \cdot {\bm \sigma}^\beta)^{p-2} \tonde{{\bf v}_2 \cdot {\bm \sigma}^\alpha} \tonde{{\bf v}_1 \cdot {\bm \sigma}^\beta}.
  \end{split}
 \end{equation}
 For what concerns the Hessians, one gets:
\begin{equation}\label{eq:HessTang}
\begin{split}
& \Big\langle  \tonde{ {\bf v}_1 \cdot  {\bm \nabla}^2 h^\alpha \cdot {\bf v}_2} \tonde{ {\bf v}_3 \cdot {\bm \nabla}^2 h^\beta \cdot {\bf v}_4} \Big\rangle
=\\
&\frac{p! ({\bm \sigma}^\alpha \cdot {\bm \sigma}^\beta)^{p-4}}{(p-4)!}({\bf v}_1 \cdot {\bm \sigma}^\beta) ({\bf v}_2\cdot {\bm \sigma}^\beta) ({\bf v}_3\cdot {\bm \sigma}^\alpha)( {\bf v}_4 \cdot {\bm \sigma}^\alpha)+\\
 &\frac{p!}{(p-3)!}({\bm \sigma}^\alpha \cdot {\bm \sigma}^\beta)^{p-3} \,{({\bf v}_1 \cdot {\bf v}_4 )( {\bf v}_2 \cdot {\bm \sigma}^\beta)( {\bf v}_3 \cdot {\bm \sigma}^\alpha)}+\\
  &\frac{p!}{(p-3)!}({\bm \sigma}^\alpha \cdot {\bm \sigma}^\beta)^{p-3} \,{({\bf v}_2 \cdot {\bf v}_4)( {\bf v}_1 \cdot {\bm \sigma}^\beta)( {\bf v}_3\cdot {\bm \sigma}^\alpha)}+\\
  &\frac{p!}{(p-3)!}({\bm \sigma}^\alpha \cdot {\bm \sigma}^\beta)^{p-3}\, {({\bf v}_1 \cdot {\bf v}_3)( {\bf v}_2 \cdot {\bm \sigma}^\beta)( {\bf v}_4 \cdot {\bm \sigma}^\alpha)}+\\
   &\frac{p!}{(p-3)!}({\bm \sigma}^\alpha \cdot {\bm \sigma}^\beta)^{p-3}\, {({\bf v}_2 \cdot {\bf v}_3)( {\bf v}_1 \cdot {\bm \sigma}^\beta)( {\bf v}_4\cdot {\bm \sigma})}+\\
 &\frac{p! ({\bm \sigma}^\alpha \cdot {\bm \sigma}^\beta)^{p-2} }{(p-2)!}\quadre{( {\bf v}_1\cdot  {\bf v}_3)( {\bf v}_2\cdot  {\bf v}_4)+( {\bf v}_1\cdot  {\bf v}_4)( {\bf v}_2\cdot  {\bf v}_3)}.
 \end{split}
\end{equation}
Finally, the correlations between Hessians and gradients read: 
\begin{equation}\label{eq:CorelationsHessianGrad}
 \begin{split}
& \Big\langle  \tonde{ {\bf v}_1 \cdot  {\bm \nabla}^2 h^\alpha \cdot {\bf v}_2} \tonde{{\bm \nabla} h^\beta \cdot  {\bf v}_3 } \Big\rangle
 =\\
 &p(p-1)(p-2) ({\bm \sigma}^\alpha \cdot {\bm \sigma}^\beta)^{p-3} ({\bf v}_1 \cdot {\bm \sigma}^\beta) ({\bf v}_2 \cdot {\bm \sigma}^\beta)  ({\bf v}_3\cdot {\bm \sigma}^\alpha)+ \\
 & p(p-1) ({\bm \sigma}^\alpha \cdot {\bm \sigma}^\beta)^{p-2} ({\bf v}_1 \cdot {\bf v}_3) ({\bf v}_2 \cdot {\bm \sigma}^\beta)+ \\
 & p(p-1) ({\bm \sigma}^\alpha \cdot {\bm \sigma}^\beta)^{p-2}({\bf v}_2 \cdot {\bf v}_3) ({\bf v}_1 \cdot {\bm \sigma}^\beta) .
 \end{split}
\end{equation}
The covariances of the components along all the directions ${\bf v}_i$ that are orthogonal to the ${\bm \sigma}^\alpha$ with $\alpha=0, \cdots, n$ have a simple form. We thus choose the bases $\mathcal{B}[{\bm \sigma}^\alpha]$ in each tangent plane in such a way that the last $n$ vectors ${\bf e}^\alpha_{N-n-1}, \cdots, {\bf e}^\alpha_{N-1}$, together with the normal direction ${\bm \sigma}^\alpha$, span the $(n+1)$-dimensional subspace $S \equiv \text{span} \grafe{{\bm \sigma}^0,{\bm \sigma}^1, \cdots, {\bm \sigma}^n}$, while the remaining $N-1-n$ vectors span the orthogonal subspace $S^\perp$. Since for each $\alpha=0,1, \cdots, n$ the vectors generating $S^\perp$ are automatically orthogonal to ${\bm \sigma}^\alpha$, they can be chosen to be equal in each tangent plane, independently of $\alpha$. We denote these vectors simply with ${\bf e}_i$ for $i=1, \cdots, N-1-n$. 
On the contrary, the $n$ vectors ${\bf e}^\alpha_{N-1-n}, \cdots, {\bf e}^\alpha_{N-n}$ have to be chosen in an $\alpha$-dependent way, since they have to be orthogonal to the normal direction ${\bm \sigma}^\alpha$. Notice that the sets $\tilde{\mathcal{B}}[{\bm \sigma}^\alpha]\equiv \grafe{\mathcal{B}[{\bm \sigma}^\alpha], {\bm \sigma}^\alpha}$ are orthonormal bases of the full $N$-dimensional space in which the sphere is embedded, which can be mapped into each others by unitary transformations. \\
The components of the gradients and Hessians along the first $M=N-1-n$ directions in each tangent plane are uncorrelated with each others, and are uncorrelated with the energy fields of all replicas. They satisfy:
\begin{equation}
 \begin{split}
  & \Big\langle \tonde{{\bm \nabla} h^\alpha \cdot {\bf e}_i} \tonde{{\bm \nabla} h^\beta \cdot {\bf e}_j} \Big\rangle =p ({\bm \sigma}^\alpha \cdot {\bm \sigma}^\beta)^{p-1} \delta_{ij},
  \end{split}
\end{equation}
and
\begin{equation}
 \begin{split}
&  \Big\langle \tonde{ {\bf e}_i \hspace{-.08 cm} \cdot \hspace{-.08 cm}  {\bm \nabla}^2 h^\alpha  \hspace{-.08 cm}\cdot  \hspace{-.08 cm}{\bf e}_j} \tonde{ {\bf e}_k \hspace{-.08 cm} \cdot  \hspace{-.08 cm}{\bm \nabla}^2 h^\beta \hspace{-.08 cm} \cdot \hspace{-.08 cm} {\bf e}_l} \Big\rangle
=
\frac{p! ({\bm \sigma}^\alpha \cdot {\bm \sigma}^\beta)^{p-2} }{(p-2)!} \times \\
&\times \tonde{\delta_{ik} \delta_{jl}+ \delta_{il} \delta_{jk}}.
  \end{split}
\end{equation}
The covariances of the remaining components along the directions ${\bf e}^a_i$ depend instead on the particular choice of these basis vectors in each tangent plane. However, since these vectors span the subspace $S$, they can be expressed as linear combinations of the ${\bm \sigma}^\alpha$, implying that their covariances are functions only of the overlaps between replicas, and can thus be parametrized by $q$ and $q_{ab}= {\bm \sigma}^{a} \cdot {\bm \sigma}^b$: the joint and conditional distributions of the rescaled energy field, its gradient and Hessian thus depend only on these parameters, implying that the action in \eqref{eq:Full3} in turns depends only on these parameters.\\
To perform explicit calculations in the following, we introduce one specific choice of these basis vectors ${\bf e}^a_i$ in each tangent plane. For the first replica ${\bm \sigma}^1$, we set:
 $$ {\bf e}_{M+k}^1=\frac{1}{\sqrt{(k+1)k (1-q_1)}} \tonde{\sum_{b=2}^{k+1} {\bm \sigma}^b - k {\bm \sigma}^{k+2}}$$ for $1 \leq k \leq n-2$, while
 \begin{equation*}
  \begin{split}
   {\bf e}_{N-2}^1&=\sqrt{\frac{n (1-q^2)}{A}}\sum_{b=2}^n {\bm \sigma}^b-\sqrt{\frac{n (1-q^2)}{A}}(n-1)q_1 {\bm \sigma}^1\\
  &-\sqrt{\frac{n}{A (1-q^2)}}{(n-1) q(1-q_1)} \tonde{q {\bm \sigma}^1-{\bm \sigma}^0}  
  \end{split}
 \end{equation*}
with the proper normalization factor  $A=n(n-1)(1-q_1) \quadre{1 - n q^2 +(n-1) q_1}$, and 
\begin{equation*}
  {\bf e}_{N-1}^1=\frac{1}{\sqrt{1-q^2}}\tonde{q {\bm \sigma}^1-{\bm \sigma}^0}.
\end{equation*}
This corresponds to choosing a unique vector, $ {\bf e}_{N-1}^1$, having non-zero overlap with the fixed point ${\bm \sigma}^0$. Analogous choices can be made for any replica $a$ with $a=2, \cdots, n$. As it will become clear in the following, this choice of bases is made to simplify the calculation of the conditional statistics of the Hessian.

\subsection{Statistics of the conditioned Hessians (I)}\label{app:ConditionedHessian}
\noindent
In this section we discuss the statistics of the $n$ Hessian matrices $\mathcal{H}^a$, conditioned to the gradients ${\bf g}^\alpha$ and to the energy fields $h^\alpha$ at the $n+1$ points ${\bm \sigma}^\alpha$. This is a necessary information to compute the joint expectation value in Eq.~\eqref{eq:Full}. We denote with $\tilde{\mathcal{H}}^a$ the matrices obeying this conditional law, and assume from now on that the overlaps between replicas satisfy the RS ansatz $q_{ab} \equiv q_1$.\\
As it follows from \eqref{eq:relHessians}, the conditioned matrix  $\tilde{\mathcal{H}}^a$ equals to 
\begin{equation}\label{eq:ShiftedHessian}
 \tilde{\mathcal{H}}^a=\tilde{\mathcal{M}}^a- \sqrt{2 N} p \epsilon \hat{1},
\end{equation}
where $\tilde{\mathcal{M}}^a$ is the Hessian projected onto the tangent plane, conditioned to gradients and energies. We aim at computing the covariances of the components $\tilde{\mathcal{M}}_{ij}^a$. We group all the independent components of the un-conditioned matrices $\mathcal{M}^a$ into an $n N(N+1)/2$-dimensional vector ${\bf M}=({\bf M}_{0}, {\bf M}_{1/2}, {\bf M}_{1})$, where ${\bf M}_\gamma= ({\bf M}_\gamma^{1}, \cdots, {\bf M}_\gamma^{n})$ for $\gamma \in \grafe{0, 1/2,1}$.
The vectors ${\bf M}_{0},  {\bf M}_{1}$ and ${\bf M}_{1/2}$ group the Hessians coordinates along directions that belong both to $S^\perp$, or both to $S$, or one to each subspace, respectively:
\begin{equation*}
 \begin{split}
 & {\bf M}_0^{a}= (\mathcal{M}_{11}^a,\mathcal{M}_{22}^a, \cdots,\mathcal{M}_{MM}^a,\mathcal{M}_{12}^a, \cdots, \cdots,\mathcal{M}_{M-1M}^a)\\
  &  {\bf M}_{1/2}^{a}= (\mathcal{M}_{1M+1}^a,\mathcal{M}_{1M+2}^a, \cdots, \cdots, \cdots,\mathcal{M}_{MN-1}^a)\\
   &   {\bf M}_1^{a}= (\mathcal{M}_{M+1M+1}^a, \cdots,\mathcal{M}_{N-1 N-1}^a,\mathcal{M}_{M+1M+2}^a,  \cdots),
 \end{split}
\end{equation*}
where $M=N-n-1$. Analogously, we define the $(n+1) N$-dimensional vector $\tilde{{\bf g}}=(\tilde{{\bf g}}_{0}, \tilde{{\bf g}}_1)$, with $\tilde{{\bf g}}_{\gamma}= (\tilde{{\bf g}}_\gamma^0,\tilde{{\bf g}}_\gamma^1, \cdots, \tilde{{\bf g}}_\gamma^n)$, and:
\begin{equation*}
 \begin{split}
\tilde{{\bf g}}_0^\alpha&= (g_1^\alpha,\cdots,g_M^\alpha),\\  
\tilde{{\bf g}}_1^\alpha&=(g_{M+1}^\alpha, \cdots, g_{N-1}^a, \tilde{g}^\alpha_N).
 \end{split}
\end{equation*}
Here $\tilde{g}_N^\alpha = {\bm \nabla}h^\alpha \cdot {\bm \sigma}^\alpha= p\, h^\alpha$, and thus conditioning to $h^a= \sqrt{2 N} \epsilon$ and $h^0= \sqrt{2 N} \epsilon_0$ is equivalent to conditioning to $\tilde{g}_N^a= \sqrt{2 N} p\; \epsilon$ and $\tilde{g}_N^0= \sqrt{2 N} p\; \epsilon_0$. \\
Before conditioning, the components in the block $\mathcal{M}^a_\gamma$ of the replica ${\bm \sigma}^a$ are correlated only with the component in the correspondent block $\mathcal{M}^b_\gamma$ of the other replicas ${\bm \sigma}^b$, since the covariance matrix  ${\bf M}$ has a block-diagonal structure:
\begin{equation*}
\begin{split}
 \hat{\Sigma}_{{\bf M}{\bf M}}&= \begin{pmatrix}
                \hat{\Sigma}_{{\bf M}{\bf M}}^{0} &&0&&0\\
                0 &&\hat{\Sigma}_{{\bf M}{\bf M}}^{1/2}&&0\\
                0 &&0&&\hat{\Sigma}_{{\bf M}{\bf M}}^{1}
              \end{pmatrix}.
               \end{split}
\end{equation*}
Since,
\begin{equation*}
\begin{split}
                             \hat{\Sigma}_{\tilde{{\bf g}}\tilde{{\bf g}}}&= \begin{pmatrix}
                \hat{\Sigma}_{\tilde{{\bf g}}\tilde{{\bf g}}}^{0} &&0\\
                             0&&\hat{\Sigma}_{\tilde{{\bf g}}\tilde{{\bf g}}}^{1}
              \end{pmatrix}
              \end{split}
\end{equation*}
and the covariances between ${\bm M}$ and $\tilde{{\bf g}}$, see \eqref{eq:CorelationsHessianGrad}, are of the form:
\begin{equation*}
  \hat{\Sigma}_{{\bf M}\tilde{{\bf g}}}= \begin{pmatrix}
                0 && 0\\
                \hat{\Sigma}_{{\bf M}\tilde{{\bf g}}}^{\frac{1}{2} 0} &&0\\
                0 &&\Sigma_{{\bf M}\tilde{{\bf g}}}^{1 1}
              \end{pmatrix},
\end{equation*}
this implies:
\begin{equation}\label{eq:CovCart}
  \hat{\Sigma}_{{\bf M}|\tilde{{\bf g}}}=
   \begingroup
\renewcommand*{\arraystretch}{1.5}
  \begin{pmatrix}
                \hat{\Sigma}_{{\bf M}{\bf M}}^{0} &\hspace{-1.5 cm}0&\hspace{-1.5 cm}0\\
                0 &\hspace{-.5 cm}\hat{\Sigma}_{{\bf M}{\bf M}}^{1/2}- \hat{\Sigma}_{{\bf M}\tilde{{\bf g}}}^{\frac{1}{2} 0}  (\hat{\Sigma}_{\tilde{{\bf g}}\tilde{{\bf g}}}^{-1})^{00}\hat{\Sigma}_{\tilde{{\bf g}}{\bf M}}^{0 \frac{1}{2}} &\hspace{-1.5 cm}0\\
                0 &\hspace{-1.5 cm}0&\hspace{-2.3 cm}\hat{\Sigma}_{{\bf M}{\bf M}}^{1}- \hat{\Sigma}_{{\bf M}\tilde{{\bf g}}}^{1 1}  (\hat{\Sigma}_{\tilde{{\bf g}}\tilde{{\bf g}}}^{-1})^{11}\hat{\Sigma}_{\tilde{{\bf g}}{\bf M}}^{11}
              \end{pmatrix}.
                \endgroup
          \end{equation}        
       Thus the conditioning to the gradients and energies preserves this block-structure of the matrix elements; moreover, the covariances of the largest blocks $\mathcal{M}^a_0$ are left untouched by the conditioning, since the subspace $S^\perp$ is blind to the presence of the other replicas. Thus, the components of this block form a GOE matrix with variance $\sigma^2=p(p-1)$. This is the only relevant information to determine the expression of the action, see the following section. To characterize the stability of the stationary points counted by the complexity, instead, it is necessary to determine the conditional distribution of the remaining components.\\
For what concerns $\hat{\Sigma}^{1/{2}}_{{\bf M}|\tilde{{\bf g}}}$, we have:
\begin{equation*}
 \tonde{\hat{\Sigma}^{1/2}_{{\bf MM}}}^{ab}_{i j, k l}=\langle \mathcal{M}^a_{i j} \mathcal{M}^b_{k l}\rangle=\delta_{i k} S^{ab}_{j l}, 
\end{equation*}
where $S^{ab}$ is a block of size $n\times n$, equal for every $i$, with components
\begin{equation}
\begin{split}
S^{ab}_{j l}&= p(p-1)(p-2)q_1^{p-3} ({\bf e}^a_j \cdot {\bm \sigma^b}) ({\bf e}^b_l \cdot {\bm \sigma^a})\\
&+p(p-1) Q_{ab}^{p-2} ({\bf e}^a_j \cdot {\bf e}^b_l).
\end{split}
\end{equation} 
Additionally, for $\beta=0, \cdots, n$ it holds
\begin{equation}
 \tonde{\hat{\Sigma}^{\frac{1}{2} 0}_{{\bf M} \tilde{{\bf g}}}}^{a \beta}_{i j, k}= \langle \mathcal{M}^a_{i j}\, { g}^\beta_{k} \rangle= \delta_{i k} p(p-1) Q_{a\beta}^{p-2} ( {\bf e}^a_j\cdot {\bm \sigma}^\beta),
 \end{equation}
and
\begin{equation*}
 (\hat{\Sigma}_{\tilde{{\bf g}}\tilde{{\bf g}}}^{0})^{-1}= \frac{1}{p}
 \left( \begin{array}{ccccc} 
 \alpha_0 \hat{1}  & \beta_0 \hat{1} & \cdots & \cdots & \beta_0\hat{1} \\
 \beta_0 \hat{1}  & \alpha_1 \hat{1} & \beta_1 \hat{1} & \cdots & \beta_1\hat{1} \\ 
 \beta_0 \hat{1} &\beta_1 \hat{1} &\alpha_1 \hat{1}&\cdots & \beta_1 \hat{1}\\ 
\cdots  &\cdots  & \cdots  &\cdots &\cdots\\ 
 \beta_0\hat{1}  & \beta_1 \hat{1} & \cdots&\cdots &\alpha_1 \hat{1} 
 \end{array} 
\right),
\end{equation*}
where the blocks have dimension $M \times M$, and
\begin{equation}\label{eq:InverseSigmaG0}
\begin{split}
  \alpha_0&=1-\frac{n q^{2p-2}}{-1+n q^{2p-2}+q_1^{p-1}-n q_1^{p-1}}\\
 \beta_0&=\frac{q^{p-1}}{-1+n q^{2p-2}+q_1^{p-1}-n q_1^{p-1}}\\
 \alpha_1&=-\frac{1-(n-1) q^{2p-2}+(n-2)q_1^{p-1}}{(1-q_1^{p-1})(-1+n q^{2p-2}+q_1^{p-1}-n q_1^{p-1})}\\
 \beta_1&=\frac{q_1^{p-1}-q^{2p-2}}{(1-q_1^{p-1})(-1+n q^{2p-2}+q_1^{p-1}-n q_1^{p-1})}.
\end{split}
\end{equation}
Doing the matrix product, we find
\begin{equation}\label{eq:CorelationsBlock1half}
\begin{split}
 &\tonde{\hat{\Sigma}^{1/2}_{{\bf M|\tilde{{\bf g}}}}}^{ab}_{i k, j l}=\delta_{i j}\; p(p-1) T^{ab}_{k l},
 \end{split}
 \end{equation}
with
\begin{equation}\label{eq:CorelationsBlock1half}
\begin{split}
 &T^{ab}_{k l}=Q_{ab}^{p-2}({\bf e}_k^a \cdot {\bf e}_l^b)+(p-2) q_1^{p-3}({\bf e}^a_k \cdot {\bm \sigma}^b)({\bm \sigma}^a \cdot {\bf e}_l^b)-\\
  &(p-1) \Big\{ \alpha_0 q^{2p-4}({\bf e}^a_k \cdot {\bm \sigma}^0)({\bm \sigma}^0 \cdot {\bf e}_l^b)+\\
  &\beta_0 (q q_1)^{p -2}\sum_{c=1}^n\quadre{({\bf e}^a_k \cdot {\bm \sigma}^c)({\bm \sigma}^0 \cdot {\bf e}_l^b)+({\bf e}^a_k \cdot {\bm \sigma}^0)({\bm \sigma}^c \cdot {\bf e}_l^b)}+\\
  &\frac{ q_1^{2p -4}}{1-q_1^{p-1}} \sum_{c (\neq a,b)=1}^n ({\bf e}^a_k \cdot {\bm \sigma}^c) ({\bf e}^b_l \cdot {\bm \sigma}^c)+\\
  &\beta_1 q_1^{2p-4} \sum_{c (\neq a)=1}^n \sum_{d (\neq b)=1}^n ({\bf e}^a_k \cdot {\bm \sigma}^c) ({\bf e}^b_l\cdot {\bm \sigma}^d)\Big\},
 \end{split}
 \end{equation}
 for $k, l= M+1, \cdots, N-1$. The averages of these  components equals to zero after the conditioning, since they are proportional to the elements of $\tilde{{\bf g}}^\alpha_0$, which are all set to zero.\\
 It remains to characterize the conditional distribution of the components ${\bf M}_{1}$. As it appears in the following, the covariances of these components do not enter in the stability analysis, while their non-zero averages induced by the conditioning do. We thus focus on the latter. Following the strategy illustrated in Ref.\cite{SpikedRepKacRice} and using the fact that, for each $a$, ${\bf e}^a_{N-1}$ is the only vector in the tangent plane at ${\bm \sigma}^a$ having non-zero overlap with ${\bm \sigma}^0$, we find:
 \begin{equation}\label{eq:Averages}
 \begin{split}
  \frac{\langle \tilde{M}^a_{ij}\rangle}{\sqrt{2 N}}&= \lambda_1 \delta_{i,j}\delta_{j,N-1} + \lambda_2 \sum_{b (\neq a)} ({\bf e}_i^a \cdot {\bm \sigma}^b)({\bf e}_j^a \cdot {\bm \sigma}^b)+\\
  &+\lambda_3 \tonde{\delta_{i, N-1} \sum_{b (\neq a)} ({\bf e}_j^a \cdot {\bm \sigma}^b)+ \delta_{j, N-1}\sum_{b (\neq a)} ({\bf e}_i^a \cdot {\bm \sigma}^b)}\\
  &+\lambda_4 \sum_{b (\neq a)} ({\bf e}_i^a \cdot {\bm \sigma}^b) \sum_{c (\neq a)} ({\bf e}_j^a \cdot {\bm \sigma}^c),
  \end{split}
 \end{equation}
where $\lambda_i$ are constants that depend explicitly on $q, q_1, \epsilon, \epsilon_0$ and $n$. Note that, with the choice of basis discussed in the previous section, it holds 
\begin{equation}\label{eq:AvDiag}
\sum_{b \neq a} ({\bf e}^a_i \cdot {\bm \sigma}^b) ({\bf e}^a_j \cdot {\bm \sigma}^b)= \delta_{ij} (1-q_1)
\end{equation}
and $\sum_{b \neq a} ({\bf e}^a_i \cdot {\bm \sigma}^b)=0$ for any $i=M+1, \cdots, N-3$. The only  non-zero averages to be determined are thus $\langle \tilde{M}^a_{ij}\rangle$ with $i, j \in \grafe{N-2, N-1}$. In the following, we will evaluate these average at the saddle point value for $q_1$, see Sec.~\ref{app:IsolatedEigenvalue}.

\subsection{Derivation of the action }\label{app:FullDerivation}
\noindent
Having characterized the statistics of the conditioned Hessians, in this section we derive the explicit expression of $\Sigma(\epsilon,\hat{Q}|\epsilon_0)$ in \eqref{eq:LinearExpansion}.\\
The expression of the moments \eqref{eq:Full} can be rewritten as 
\begin{equation}\label{eq:Full3bis}
M_n=\hspace{-0.1 cm}  \int \hspace{-0.15 cm}\prod_{a < b=1}^n \hspace{-0.1 cm} dQ_{ab} \, V_n(\hat{Q})\,
 \mathcal{E}_n(\epsilon, \hat{Q}) \,P_{n}(\epsilon, \hat{Q}| \epsilon_0),
\end{equation}
where $\mathcal{E}_n$ and $P_{n}$ are the expectation value and the joint distribution in Eq.\eqref{eq:Full}, now expressed as a function of the $(n+1) \times (n+1)$ overlap matrix $\hat{Q}$, while $V_n$ is an entropic contribution reading:  
\begin{equation}\label{eq:VolGeneral}
 V_n(\hat{Q})\hspace{-0.1 cm}=\hspace{-0.1 cm}\int \hspace{-0.1 cm}\prod_{a=1}^n d{\bm \sigma}^a \delta \tonde{{\bm \sigma}^a\hspace{-0.1 cm} \cdot\hspace{-0.08 cm} {\bm \sigma}^0- q}  \prod_{a \leq b=1}^n \delta \tonde{Q_{ab}-{\bm \sigma}^{a}\hspace{-0.1 cm}\cdot \hspace{-0.08 cm}{\bm \sigma^{b}}}.\\
\end{equation}
The explicit form of $\Sigma(\epsilon,\hat{Q}| \epsilon_0)$ is obtained extracting the leading order contribution in $N n$ of each of the three terms in \eqref{eq:Full3bis}. We consider each of them separately in the following three subsections, and collect all terms in the final expression, Eq.\ref{eq:FinalComplexityPreSaddle}. The calculation is done under the assumption that $q_{ab} \equiv q_1$.

\subsubsection{Phase space term}
\noindent
The calculation of the phase-space term is standard, see also Ref. \cite{SpikedRepKacRice}, and leads to:
\begin{equation}\label{eq:Volume}
 V_n= e^{\frac{N n}{2} \quadre{\log \tonde{\frac{2 \pi e (1-q_1)}{N}} + \frac{q_1- q^2}{1-q_1}} + o(Nn)}.
\end{equation} 
Note that, for $q_1=q^2$, the second term at the exponent vanishes, and the expression \eqref{eq:Volume} reproduces the form of \eqref{eq:VolGeneral} for $n=1$, which is the term obtained when performing the annealed calculation of the complexity.

\subsubsection{Joint distribution of energies and gradients}
\noindent
The joint distribution of the gradients and energies of the $n$ replicas ${\bm \sigma}^a$ conditioned to ${\bm \sigma}^0$ can be obtained as 
\begin{equation}
 P_{n}(\epsilon,\hat{Q}| \epsilon_0)=\frac{P_{n+1}(\epsilon,\epsilon_0,\hat{Q})}{P_{1}(\epsilon_0, \hat{Q})},
\end{equation}
where $P_{n+1}(\epsilon,\epsilon_0,\hat{Q})$ is the joint distribution of the gradients and energies of the $n+1$ points ${\bm \sigma}^\alpha$, $\alpha=0, 1, \cdots, n$, evaluated at ${\bf g}^\alpha= {\bf 0}$, $h^a= \sqrt{2 N} \epsilon$ and ${h^0}= \sqrt{2 N} \epsilon_0$, while $p_{1}(\epsilon_0, \hat{Q})$ is the density of the gradient and energy field of ${\bm \sigma}^0$, evaluated at ${\bf g}^0= {\bf 0}$ and $h^0= \sqrt{2N}\epsilon_0$. From the fact that the gradient and energy field at the same point ${\bm \sigma}$ are uncorrelated, see \eqref{eq:AvGrad}, it follows that:
\begin{equation}
P_1(\epsilon_0, \hat{Q})= \frac{e^{- N \epsilon_0^2}}{\sqrt{2 \pi}} \frac{1}{(2 \pi p)^{(N-1)/2}}= e^{- N \tonde{\epsilon_0^2 + \frac{\log (2 \pi p)}{2}} + o(N)}.
\end{equation}
To compute $P_{n+1}(\epsilon,\epsilon_0,\hat{Q})$, it is convenient to proceed as in Ref.~\cite{SpikedRepKacRice} and first determine the joint distribution of the $N$-dimensional vectors ${\bf \tilde{g}}\quadre{{\bm \sigma}^\alpha}\equiv{\bf \tilde{g}}^\alpha= (\tilde{g}_0^\alpha, g^\alpha_{1}, g^\alpha_{2}, \cdots, g^\alpha_{N-1})$, whose last $N-1$ components are the components of the gradient ${\bm \nabla} h^\alpha$ in the chosen basis $\mathcal{B}[{\bm \sigma}^\alpha]$ of the tangent plane at ${\bm \sigma}^\alpha$, $g^\alpha_{\beta}= {\bm \nabla} h[{\bm \sigma}^\alpha] \cdot {\bf e}^\alpha_\beta$, while the first component is proportional to the energy field, $\tilde{g}_0^\alpha ={\bm \nabla} h[{\bm \sigma}^\alpha] \cdot {\bm \sigma}^\alpha= p\; h[{\bm \sigma}^\alpha]$. The joint density of the vectors  ${\bf \tilde{g}}^\alpha$ evaluated at $\tilde{g}_i^\alpha=0$ for $i=1, \cdots, N-1$ equals to:
\begin{equation}\label{eq:GaussianEuclideanGrad}
 P\tonde{\grafe{\tilde{g}_0^\alpha, {\bf 0}}_{\alpha=0}^n}= \frac{\text{exp}\grafe{-\frac{1}{2} \sum_{\alpha,\beta=0}^n \tilde{g}_0^\alpha {\bm \sigma}^\alpha \cdot [{\hat{C}}^{-1}]^{\alpha \beta}\cdot {\bm \sigma}^\beta  \tilde{g}_0^\beta}}{(2 \pi)^{\frac{(n+1) N}{2}} |\text{det} \;  \hat{C}|^{\frac{1}{2}}},
\end{equation}
where $\hat{C}$ is the covariance matrix of the gradients in the reference frame of the extended $N$-dimensional space, 
\begin{equation}
 {C}_{ij}^{\alpha \beta}\equiv \langle {\bm \nabla} h^\alpha_i {\bm \nabla} h^\beta_j \rangle = p Q_{\alpha \beta}^{p-1} \delta_{ij}+ p(p-1)Q_{\alpha \beta}^{p-2} \sigma^\alpha_j \sigma^\beta_i.
\end{equation}
Performing the change of variables at the exponent, we obtain
\begin{equation}\label{eq:JointDensity2}
 P_{n+1}(\epsilon,\epsilon_0,\hat{Q})= \frac{p^{2n +2}}{(2 \pi)^{\frac{(n+1) N}{2}} |\text{det} \;  \hat{C}|^{\frac{1}{2}}}e^{-N p^2 f(\epsilon, \epsilon_0,q_1, q)},
\end{equation}
where 
\begin{equation}\label{eq:quadForm}
  f(\epsilon, \epsilon_0,q_1, q)= \epsilon^2 \sum_{a, b=1}^n M^{ab}+ \epsilon_0^2 M^{00} + \epsilon_0 \epsilon \sum_{a=1}^n \tonde{M^{a0}+ M^{0a}}
\end{equation}
and 
\begin{equation}
 M^{\alpha \beta} \equiv ({\bm \sigma}^\alpha)^T \cdot [\hat{C}^{-1}]^{\alpha \beta} \cdot  {\bm \sigma}^\beta. 
\end{equation}
The contribution of the determinant in \eqref{eq:JointDensity2} is easily obtained from the fact that $\hat{C}^{\alpha \beta}=\text{diag}(\hat{A}^{\alpha \beta}, {\hat{B}}^{\alpha \beta})$, where $\hat{A}^{\alpha \beta}$ is the $(N-n-1)\times(N-n-1)$ block which gives the covariances between the gradients components in $S^\perp$, $
 \hat{A}_{ij}^{\alpha \beta}= p \delta_{ij}\grafe{ \delta_{\alpha \beta} + (1-\delta_{\alpha \beta})[q_1^{p-1}+ (\delta_{\alpha 0}+ \delta_{\beta 0}) (q^{p-1}-q_1^{p-1})]},$
 while ${\hat{B}^{\alpha \beta}}$ are $(n+1) \times (n+1)$ blocks whose elements are the covariances of the gradients components in $S$. To leading order in $N$ only the block $\hat{A}$ contributes, giving: 
\begin{equation}\label{eq:DeterLow}
\begin{split}
 &|\text{det} \; \hat{C}|=p^N \, e^{N n \tonde{\log \quadre{p(1-q_1^{p-1})} + \frac{q_1^{p-1}- q^{2p -2}}{1-q_1^{p-1}}}+ o(Nn)}.\\
 \end{split}
\end{equation}
To compute the quadratic form \eqref{eq:quadForm}, it is convenient to introduce the set of $N (n+1)$-dimensional vectors:
\begin{equation}
 \begin{split}
 \vec{{\bm \xi}}_1&=\tonde{{\bm \sigma}^0,{\bf 0}, \cdots, {\bf 0} },\\
   \vec{{\bm \xi}}_2&=\tonde{\sum_{a=1}^n{\bm \sigma}^a,{\bf 0}, \cdots, {\bf 0} },\\
 \vec{{\bm \xi}}_3&= \tonde{{\bf 0}, {\bm \sigma}^1, \cdots, {\bm \sigma}^n},\\
 \vec{{\bm \xi}}_4&= \tonde{{\bf 0}, {\bm \sigma}^0, \cdots, {\bm \sigma}^0},\\
  \vec{{\bm \xi}}_5&= \tonde{{\bf 0},\sum_{a \neq \grafe{0,1}}{\bm \sigma}^a, \cdots, \sum_{a \neq \grafe{0,n}}{\bm \sigma}^a},
 \end{split}
\end{equation}
which form a close set under the action of the matrix $\hat{C}^{-1}$. To show that this is the case, we split the covariance matrix into its diagonal $\hat{D}$ and off-diagonal $\hat{O}$ parts in the space of replicas, $\hat{C}=p \tonde{\hat{O}+\hat{D}}$, and write:
\begin{equation}
 \begin{split}
  \hat{C}^{-1}= p^{-1} \hat{D}^{-1} \tonde{\hat{1}+ \hat{O} \hat{D}^{-1}}^{-1}
 \end{split}
\end{equation}
where 
\begin{equation}
[\hat{D}^{-1}]^{\alpha \beta}_{ij}= \delta_{\alpha \beta}\tonde{\delta_{ij} -\frac{p-1}{p} \sigma^\alpha_i \sigma^\alpha_j},
\end{equation}
and
\begin{equation*}\label{eq:BlocksOp}
\begin{split}
 &[\hat{O} \hat{D}^{-1}]^{\alpha \beta}_{ij}=\\
 &(1-\delta_{\alpha \beta}) (\delta_{\alpha 0}+\delta_{\beta 0})\quadre{A' \delta_{ij}+ B' \sigma^\beta_i \sigma^\alpha_j-C' \sigma^\beta_i \sigma^\beta_j}+\\
 &(1-\delta_{\alpha \beta}) (1- \delta_{\alpha 0}-\delta_{\beta 0})\quadre{A \delta_{ij}+ B \sigma^\beta_i \sigma^\alpha_j-C \sigma^\beta_i \sigma^\beta_j},
 \end{split}
\end{equation*}
with $A'=q^{p-1}, B'=(p-1)q^{p-2}, C'=(p-1) q^{p-1}$ and $A=q_1^{p-1}, B=(p-1)q_1^{p-2}, C=(p-1) q_1^{p-1}$. Then it is immediate to show that:
\begin{equation*}\label{eq:ClosedAction}
 \begin{split}
\hat{O} \hat{D}^{-1}\vec{{\bm \xi}}_1&= A' \vec{{\bm \xi}}_4, \\
\hat{O} \hat{D}^{-1}\vec{{\bm \xi}}_2&= A' \vec{{\bm \xi}}_3+ \quadre{B'-q C' +(n-1)(B' q_1-C' q)}  \vec{{\bm \xi}}_4\\
&+A' \vec{{\bm \xi}}_5,\\
\hat{O} \hat{D}^{-1}\vec{{\bm \xi}}_3&= A' \vec{{\bm \xi}}_2 + A \vec{{\bm \xi}}_5,\\
\hat{O} \hat{D}^{-1}\vec{{\bm \xi}}_4&= A' \vec{{\bm \xi}}_1+ (B' -C' q) \vec{{\bm \xi}}_2+ A(n-1) \vec{{\bm \xi}}_4\\
&+(B-C)q  \vec{{\bm \xi}}_5, \\
\hat{O} \hat{D}^{-1}\vec{{\bm \xi}}_5&= (n-1)\quadre{A' + B' q- C' q_1}  \vec{{\bm \xi}}_2+ A(n-1) \vec{{\bm \xi}}_3\\
&+ \quadre{A(n-2)+ B-C q_1+(n-2)q_1(B -C)} \vec{{\bm \xi}}_5.
 \end{split}
\end{equation*}
To compute the action of $ \tonde{\hat{1}+ \hat{O} \hat{D}^{-1}}^{-1}$ in this closed subspace, we introduce an orthonormal basis for it given by the vectors: 
\begin{equation}\label{eq:BasisVecPlanted}
 \begin{split}
  &\vec{{\bm \chi}}_1=\vec{{\bm \xi}}_1\\
 & \vec{{\bm \chi}}_2= \frac{1}{\sqrt{n(1-n q^2 +(n-1)q_1)}}\tonde{-n q \vec{{\bm \xi}}_1+\vec{{\bm \xi}}_2}\\
 &  \vec{{\bm \chi}}_3=\frac{1}{\sqrt{n}}\vec{{\bm \xi}}_3\\
 &  \vec{{\bm \chi}}_4= \frac{1}{\sqrt{n(1-q^2)}}\tonde{- q \vec{{\bm \xi}}_3+\vec{{\bm \xi}}_4}\\
 &  \vec{{\bm \chi}}_5= \frac{1}{\sqrt{n(n-1)(1-q^2)(1-q_1)(1-n q^2 +(n-1)q_1)}}\times\\
   &\times\tonde{(n-1)(q^2-q_1) \vec{{\bm \xi}}_3- (n-1)q(1-q_1)\vec{{\bm \xi}}_4+ (1-q^2)\vec{{\bm \xi}}_5}.
 \end{split}
\end{equation}
In this basis, the action of the operator $\hat{1}+ \hat{O} \hat{D}^{-1}$ is given by the following matrix:
\begin{equation*}
\hat{1}+ \hat{O} \hat{D}^{-1}=\left[
\begin{array}{c|c }
\stackrel{2 \times 2}{\hat{1}} & \quad \stackrel{2 \times 3}{L_1} \quad \\
\hline\\
 \stackrel{3 \times 2}{L_2}&  \quad \stackrel{3 \times 3}{\hat{1}}+\stackrel{3\times 3}{L_3}
\end{array}\right]
 \end{equation*}
with blocks
 \begin{equation*}
 \begin{split}
   L_1=&
  \begin{pmatrix}
\sqrt{n} q^p& p q^{p-1} \sqrt{n - n q^2}& 0\\
   q^{p-1} S_1 & \frac{S_1 q^{p-2} \quadre{p (1-q^2)-1}}{\sqrt{1 - q^2}}& q^{p-1}\sqrt{\frac{(n-1) (1 - q_1) }{1 - q^2}}
  \end{pmatrix},\\
   L_2=&
  \begin{pmatrix}
   \sqrt{n} q^p& p q^{p-1} S_1\\
   q^{p-1} \sqrt{n - n q^2}&\frac{S_1 q^{p-2}\quadre{p (1-q^2)-1}}{\sqrt{1 - q^2}}\\
   0&q^{p-1}\sqrt{\frac{(n-1) (1 - q_1) }{1 - q^2}}
  \end{pmatrix},
  \end{split}
  \end{equation*}
  and 
\begin{equation*}
   L_3\hspace{-.08 cm}=\hspace{-.08 cm}
   \begingroup % keep the change local
\setlength\arraycolsep{.1pt}
  \begin{pmatrix}
(n-1) q_1^
  p& p \frac{(n-1)q (1 - q_1) q_1^{p-1}}{\sqrt{1 - q^2}}& 
    \frac{S_2 p q_1^{p-1}}{\sqrt{
   1-q^2}}\\
       \frac{(n-1)q (1 - q_1) q_1^{p-1}}{\sqrt{1 - q^2}}& S_3
   &
   \frac{q  q_1^{p-2} \quadre{p (1 - q_1)-1} S_2}{\sqrt{1 -q^2}}\\
  
  \frac{S_2 q_1^{p-1}}{\sqrt{
   1-q^2}}&
   \frac{S_2 q  q_1^{p-2} \quadre{p (1 - q_1)-1}}{\sqrt{1 -q^2}}& \frac{q_1^{p-2}S_4}{-1 + q^2}.
      \end{pmatrix}
       \endgroup
\end{equation*}
where
\begin{equation*}
 \begin{split}
  S_1&=\sqrt{1 -  n q^2 +(n-1) q_1},\\
  S_2&=\sqrt{(n-1) (1 - q_1) (1 - n q^2 + (n-1)q_1)},\\
  S_3&=\frac{(n-1) q_1^{p-2}(q^2 (p(q_1-1)^2-1) + q_1)}{1 - 
   q^2},\\
  S_4&=1 - n q^2 + (n - 1) q^2 q_1 - p (1-q_1) (1 - q_1 + n (q_1-q^2)).
 \end{split}
\end{equation*}
Setting
\begin{equation}
 \hat{Y} \equiv \tonde{\hat{1}+ \hat{O} \hat{D}^{-1}}^{-1}
\end{equation}
for the inverse of this matrix, we get that the quadratic form in \eqref{eq:quadForm} can be written in terms of its matrix elements in the basis \eqref{eq:BasisVecPlanted}, as 
\begin{equation}\label{eq:FormLinear}
 p^2 f(\epsilon, \epsilon_0, q_1, q)= \epsilon_0^2 Y_{11}+ \epsilon_0 \epsilon \sqrt{n}  \quadre{Y_{13}+ Y_{31}}+ \epsilon^2 n Y_{33},
\end{equation}
where the $Y_{ij}$ depend on $q, q_1$ and $n$.\\
The expression for the $Y_{ij}$ for general $n$ is rather cumbersome. A major simplification occurs for $n \to 1$, where only one replica is present. In this case the dependence on the overlap $q_1$ naturally drops, and one gets:
\begin{equation}\label{eq:Constantsn1}
 \begin{split}
&Y_{11}\to \frac{q^4 - q^{2 p} [1 + p (p-2+(3-2 p) q^2 + (p-1) q^4)]}{Y} \\ 
 & Y_{13}+ Y_{31}\to \frac{-2 q^{p+4} + 2 q^{3 p} (1 -p (1-q^2))}{Y}\\
 & Y_{33}  \to \frac{q^4 - q^{2 p} (1 + p (p-2+ (3-2 p) q^2 + (p-1) q^4))}{Y}
 \end{split}
\end{equation}
with the denominator being equal to
\begin{equation}
 Y=q^4 + q^{4 p}- 
 q^{2 p} [(p-1)^2 - 2 (p-2) p q^2 + (p-1)^2 q^4].
\end{equation}
The joint density of gradients and energies of the two stationary points ${\bm \sigma}^0$ and ${\bm \sigma}^1$ is in this case is equal to
\begin{equation}
 P_2(\epsilon, \epsilon_0, q)= \tonde{\frac{e^{- \tonde{\epsilon_0^2 Y_{11} + \epsilon_0 \epsilon (Y_{13}+ Y_{31})+ \epsilon^2 Y_{33}}}}{2 \pi p \sqrt{1- q^{2p-2}}}}^N e^{o(N)}.
\end{equation}
This is the contribution that one gets from the annealed calculation of the complexity, which, as we show in the following, is reproduced by the quenched calculation evaluated at the saddle point for $q_1$.\\
To compute the contribution to the quenched complexity, we consider the expansion of the matrix elements of $\hat{Y}$ to linear order in $n$:
\begin{equation}
\begin{split}
 &Y_{11}= 1+ n \frac{y_{11}(q_1,q)}{y(q_1,q)}+ o(n),\\
 &\sqrt{n}  \quadre{Y_{13}+ Y_{31}}= n \frac{y_{13}(q_1,q)}{y(q_1,q)}+ o(n),\\
 &n Y_{33}= n \frac{y_{33}(q_1,q)}{y(q_1,q)}+o(n)
 \end{split}
\end{equation}
with 

\begin{equation}\label{eq:MatrixElementsLinearized}
\begin{split}
y_{11}&= -p^2 q^{2p+2} (1-q_1)^2 q_1^{p+1}\\
&+q^{2 p}q_1^3(1-q_1^{p-2}) \left(q^4
   (q_1^{p-1}-1)+(1-q_1) q^{2 p}\right)\\
   &-p q^{2 p} q_1^3\left(1-q_1^{p-2}\right)\times\\
   &\times\left[(1-q_1) q^{2
   p}-q^4 \left(1-q_1^{p-1}\right)+q^2 \left(1-q_1^p\right)\right],\\
     y_{13}&=2 q^p [-q_1^2 + q_1^p (1 - p (1 - q_1))]\times\\
  &\times[-(p-1) q^{2 p} (1 -q_1) q_1 + q^4 q_1 (q_1^{p-1}-1)],\\
 y_{33}&=p^2 q^{2 + 2 p}(1 -q_1) q_1^3 (q_1^{p-1}-1)\\
 &+(p-1)^2 q^{2p} (1 -q_1) q_1 [q_1^2 + q_1^p(p (q_1-1)^2-1)]\\
  &+ q^4 q_1 (q_1^{p-1}-1)[q_1^2 + q_1^p(p(1-q_1)^2-1)] ,\\
  y&=-p(p-1)  q^{2 p+2}(1-q_1)q_1^3
   (1-q_1^{p-1})(1-q_1^{p-2})\\
   &+(p-1)^2 (1-q_1) q_1 q^{2 p}\times\\
   &\times\left(q_1^{2 p}+\left(p
   (1-q_1)^2-q_1^2-1\right) q_1^p+q_1^2\right)+\\
   &q^4 \left(q_1^p-q_1\right) \left(q_1^{2 p}+\left(p
   (1-q_1)^2-q_1^2-1\right) q_1^p+q_1^2\right).
     \end{split}
\end{equation}
Combining \eqref{eq:DeterLow} and \eqref{eq:FormLinear} we get:
\begin{equation}\label{eq:ensityGradients}
 P_{n}(\epsilon, \hat{Q}| \epsilon_0)= \frac{e^{-\frac{Nn}{2}\quadre{F(\epsilon, \epsilon_0) +  \frac{q_1^{p-1}- q^{2p-2}}{1-q_1^{p-1}}}+o(Nn)}}{[2 \pi p (1-q_1^{p-1}) ]
^{\frac{Nn}{2}}},
\end{equation}
where the linearized quadratic form is given by:
\begin{equation}
 F(\epsilon, \epsilon_0)=\frac{2}{y}\quadre{\epsilon_0^2 y_{11}+ \epsilon_0 \epsilon y_{13}+ \epsilon^2 y_{33}}.
\end{equation}

\subsubsection{Expectation value of the determinants}
\noindent
The expectation value $\mathcal{E}_n(\epsilon, \hat{Q})$ is over the joint distribution of the Hessians of the $n$ replicas ${\bm \sigma}^a$, conditioned to the values of the gradients and energy fields of all the $n+1$ points ${\bm \sigma}^\alpha$. Following exactly the same steps as in Ref.~\cite{SpikedRepKacRice}, we can argue that:
\begin{itemize}
 \item[(i)] even though the conditioned Hessian matrices $\tilde{\mathcal{H}}^a$ associated to different replicas are correlated with each others, these correlations are irrelevant when computing the leading-order term in $N$ of $\mathcal{E}_n(\epsilon, \hat{Q})$, as it holds:
 \begin{equation}\label{eq:Factorization}
  \mathcal{E}_n= \left\langle {\prod_{a=1}^n |\text{det} \tilde{\mathcal{H}}^a|} \right\rangle = e^{ \sum_{a=1}^n \langle |\text{det} \tilde{\mathcal{H}}^a| \rangle +o(N)}.
 \end{equation}
 The reason for this equality (valid at leading exponential order in $N$) is that the joint probability measure on the eigenvalue densities has the form of a large deviation principle in $e^{N^2}$. In consequence, the average
 above does not bias the measure at leading exponential order in N and one can replace the average of the exponential with the exponential of the average. See Ref.~\cite{SpikedRepKacRice} for a detailed explanation. 
 Given the equivalence between replicas, Eq.~\eqref{eq:Factorization} can be written as:
 \begin{equation}\label{eq:DensityEq}
    \mathcal{E}_n=N^{\frac{Nn}{2}} e^{Nn \int d\lambda \rho_{\text{sp}}(\lambda) \log |\lambda| + o(Nn)},
 \end{equation}
where $\rho_{\text{sp}}(\lambda)$ is the density of states of the matrices $\tilde{\mathcal{H}}^a/\sqrt{N}$. 
 \item[(ii)] The exponent \eqref{eq:DensityEq} is, to leading order in $N$, determined by the bulk of the density of states $\rho_{\text{sp}}(\lambda)$. This is governed by the largest $(N-n-1) \times (N-n-1)$ block of the Hessian, whose components are iid Gaussian variables with variance $\sigma^2=p(p-1)$ and non-zero average along the diagonal, due to the shift in \eqref{eq:ShiftedHessian}. As a result, up to subleading corrections in $1/N$ it holds:
 \begin{equation}
  \rho_{\text{sp}}(\lambda)= \frac{\sqrt{4p(p-1)- (\lambda + \sqrt{2}p \epsilon)^2}}{2 \pi p(p-1)}.
 \end{equation}
\end{itemize}
Combining these two results, we obtain
\begin{equation}\label{eq:FinalDeterminant}
 \mathcal{E}_n(\epsilon, \hat{Q})= e^{\frac{Nn}{2}  \quadre{\log N + \log [2 p (p-1)] + 2 I\tonde{\sqrt{\frac{p}{p-1}}\epsilon}}+ o(Nn)},
\end{equation}
where $I(y)=I(-y)$ is given by:
\begin{equation*}\label{eq:Inty}
\begin{split}
 I=\begin{cases} 
 \frac{y^2-1}{2} + \frac{y}{2} \sqrt{y^2-2}+ \log \tonde{\hspace{-0.05 cm} \frac{\hspace{-0.05 cm}-y+\hspace{-0.05 cm} \sqrt{y^2-2}}{2}\hspace{-0.05 cm}}  \hspace{-0.05 cm} \text{ if        } y \leq -\sqrt{2},\\
  \frac{1}{2} y^2 - \frac{1}{2} \tonde{1 + \log 2}  \text{  if        }-\sqrt{2}\leq y\leq 0.
 \end{cases}
 \end{split}
 \end{equation*}
 Because of the factorization in (\ref{eq:Factorization}), this contribution is independent on the overlap $q_1$ between replicas, and it is equal to the contribution one would get from the annealed calculation, elevated to the power $n$.

\subsection{Saddle point of the action and equivalence to annealed}\label{app:SaddlePoint}
\begin{figure}[!htbp]
\includegraphics[width=.95\columnwidth]{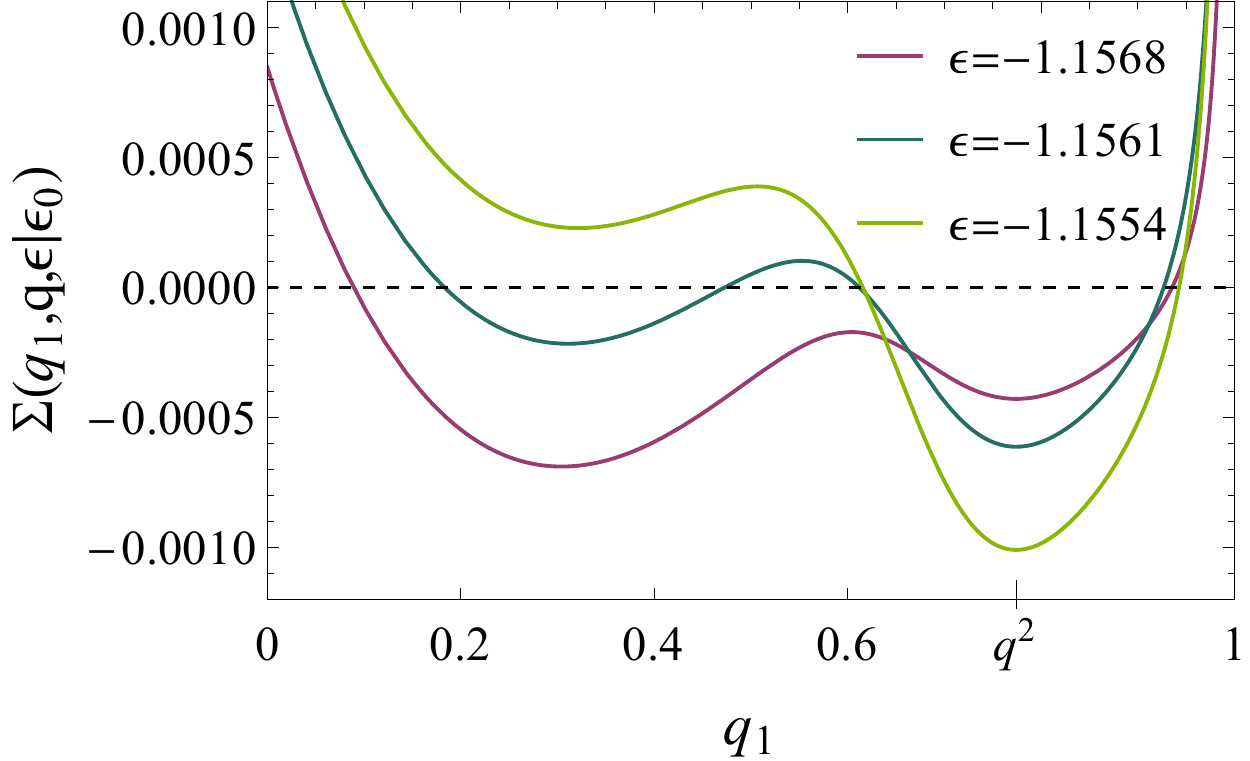}
    \caption{Complexity as a function of the overlap $q_1$ for $q=0.88$ and $\epsilon_0=-1.1582$ and different values of $\epsilon$.}\label{fig:PlotDueMinimi}
 \end{figure}

\noindent
Combining the results \eqref{eq:Volume}, \eqref{eq:ensityGradients} and \eqref{eq:FinalDeterminant}, we get that the linear order term in \eqref{eq:LinearExpansion} reads:
\begin{equation}\label{eq:FinalComplexityPreSaddle}
\begin{split}
\Sigma(\epsilon,q,q_1| \epsilon_0) 
&=\frac{1}{2} \begin{cases}\Sigma_{<}(\epsilon,q,q_1| \epsilon_0) &\mbox{ if }\epsilon \leq \epsilon_{\text{th}}(p) \\
                 \Sigma_{>}(\epsilon, q,q_1|\epsilon_0) &\mbox{ if }\epsilon > \epsilon_{\text{th}}(p)
              \end{cases}
\end{split}
\end{equation}
where $\epsilon_{\text{th}}(p)= -\sqrt{2(p-1)/p}$ and
\begin{equation}
 \begin{split}
  \Sigma_{<}&=                \log\tonde{\frac{p}{2}} + \frac{p}{p-1}\tonde{\epsilon^2+ \epsilon \tilde{z}}+2 \log \tonde{-\epsilon + \tilde{z}}+ Q,\\
                \Sigma_{>}&= \log (p-1)+ \frac{p}{p-1}\epsilon^2 + Q,
                 \end{split}
\end{equation}
with $\tilde{z}=\sqrt{\epsilon^2- \epsilon_{\text{th}}^2}$ and where $Q$ is the only term depending explicitly on $q_1$,
\begin{equation}\label{eq:ToMinimize}
 Q= \log \tonde{\frac{1-q_1}{1-q_1^{p-1}}}+\frac{q_1-q^2}{1-q_1}+ \frac{q^{2p-2}- q_1^{p-1}}{1-q_1^{p-1}}- F(\epsilon, \epsilon_0).
\end{equation}
The saddle point value for $q_1$ is therefore determined by the equation $\partial Q / \partial q_1=0$; if multiple solutions are present, the global minimum should be selected. We find that, irrespectively of the values of $\epsilon$ and $\epsilon_0$, $q_1=q^2$ is always a solution to this equation. For $q$ sufficiently large, a second minimum appears, which for certain $\epsilon$ is the deepest one, see Fig. \ref{fig:PlotDueMinimi}. However, when this happens the corresponding complexity is found to be always smaller than zero, corresponding to the absence of stationary points.  
In conclusion, we find that the relevant saddle point solutions for $q_1$ is $q_1=q^2$. This has a simple geometrical interpretation: $q^2$ is the minimal possible overlap between vectors on the sphere that are constrained to be at fixed overlap $q$ with a fixed direction; it corresponds to the $n$ replicas having zero overlap with each others in the  subspace orthogonal to the direction singled out by ${\bm \sigma}^0$. When plugging this value into \eqref{eq:FinalComplexityPreSaddle}, we find that the quenched complexity reproduces the annealed one, which is obtained taking the logarithm of the average number of stationary points at fixed overlap with a minimum ${\bm \sigma}^0$. In particular, \eqref{eq:ToMinimize} reduces to:
\begin{equation}
 Q \to \log \tonde{\frac{1-q^2}{1-q^{2p-2}}}- \tonde{\epsilon_0^2 U_0(q)+ \epsilon_0 \epsilon U(q)+ \epsilon^2 U_1(q)},
\end{equation}
with
\begin{equation}
 \begin{split}
  U_0(q)&=\frac{q^{2 p} \left(-q^{2 p}+p \left(q^2-q^4\right)+q^4\right)}{q^{4 p}-\left((p-1)^2 (1+q^4)-2 (p-2) p q^2\right) q^{2 p}+q^4},\\
  U(q)&=\frac{2 q^{3 p} \left(p \left(q^2-1\right)+1\right)-2 q^{p+4}}{q^{4 p}-\left((p-1)^2 (1+q^4)-2 (p-2) p q^2\right) q^{2 p}+q^4},\\
  U_1(q)&=\frac{q^4-q^{2 p} \left(p \left((p-1) q^4+(3-2 p) q^2+p-2\right)+1\right)}{q^{4 p}-\left((p-1)^2 (1+q^4)-2 (p-2) p q^2\right) q^{2 p}+q^4}.
 \end{split}
\end{equation}
These expressions reproduce the limit of \eqref{eq:FormLinear} when $n \to 1$, and thus are equally obtained when performing the annealed calculation for the complexity (see e.g. \cite{subag}). 

\subsection{Statistics of the conditioned Hessian (II): at the saddle point}\label{app:IsolatedEigenvalue}
\noindent
 \subsubsection{Variances and averages at the saddle point}
 \noindent
We now discuss the statistics of the Hessian matrices, evaluated at the saddle point value for $q_1$. Setting $q_1=q^2$, we find from \eqref{eq:InverseSigmaG0} that $\beta_1=0$, and that \eqref{eq:CorelationsBlock1half} for $a=b$ reduces to
\begin{equation}
\begin{split}
 T_{kl}^{aa}&=\delta_{kl} \grafe{1- c(q)\quadre{\delta_{k N-1}+ (1-\delta_{k N-1}) q^{2p-4}}}
 \end{split}
\end{equation}
with
\begin{equation}
 c(q)=\frac{(p-1)(1-q^2) q^{2p-4}}{1-q^{2p-2}}.
\end{equation}
For what concerns the averages in \eqref{eq:Averages}, we find instead that for $q_1=q^2$, the constant $\lambda_4$ vanishes, while 
$\sum_{b (\neq a)} ({\bf e}^a_{N-2} \cdot {\bf \sigma}^b)^2= (1-q^2)$. Therefore, in this limit the average $\langle \tilde{M}^a_{N-2 \, N-2}\rangle / \sqrt{N}$ becomes equal to the ones of the components $\langle \tilde{M}^a_{i \, i}\rangle / \sqrt{N} $ for $i=M+1, \cdots, N-3$, see \eqref{eq:AvDiag}, and it is given by:
\begin{equation}\label{eq:Nu}
 \nu \equiv \sqrt{2}\kappa_2(q, \epsilon, \epsilon_0) (1-q^2),
\end{equation}
where $\kappa_2(q, \epsilon, \epsilon_0)= \lim_{q_1 \to q^2} \lambda_2(q,q_1, \epsilon, \epsilon_0, n)$ is independent on $n$ and reads explicitly:
\begin{equation}
 \kappa_2(q, \epsilon, \epsilon_0)=\frac{p (p-1) \kappa_2^{(n)}(q, q_1, \epsilon, \epsilon_0) }{\kappa_2^{(d)}(q, q_1, \epsilon, \epsilon_0)}
\end{equation}
and
\begin{equation*}
 \begin{split}
  \kappa_2^{(d)}=&q^4 \left(q^{4-2 p}+q^{2 p}-(p-1)^2 (1+q^4)+2 (p-2) p q^2\right)\\
   \kappa_2^{(n)}=& \epsilon \quadre{\left((p-2) q^2-p+1\right) q^{2 p}+q^4}+\\
   &\epsilon_0 \quadre{\left((p-1) q^2-p+2\right) q^{p+2}-q^{3 p}}.
 \end{split}
\end{equation*}
Moreover, for $q_1= q^2$ we find $\lambda_3- q \sqrt{1-q^2}\lambda_2 = 0$, implying that $\langle \tilde{M}^a_{N-1 \, N-2} \rangle = 0$. The remaining non-zero average to be computed is $\langle \tilde{M}^a_{N-1 \, N-1} \rangle/ \sqrt{N} =\sqrt{2} [\lambda_1 + (n-1)q \sqrt{1-q^2} \tonde{q \sqrt{1-q^2}\lambda_2-2 \lambda_3}]$, which at the saddle point is $n$-independent and explicitly equals to:
\begin{equation}\label{eq:Mu}
 \begin{split}
 \mu(q, \epsilon, \epsilon_0) \equiv\frac{\sqrt{2} (p-1) p \left(1-q^2\right) \left(a_0(q) \epsilon_0-a_1(q) \epsilon \right)}{a_2(q)}
 \end{split}
\end{equation} 
with 
\begin{equation}
\begin{split}
 a_1&=q^{3 p}+ q^{p+2}\left(p-2-(p-1) q^2\right)\\
 a_0&= q^4+ q^{2 p}\left(1-p +(p-2)q^2\right)\\
 a_2&=q^{6-p}+q^{3 p+2}- q^{p+2}\left((p-1)^2 (q^4+1)-2 (p-2) p q^2\right).
   \end{split}
\end{equation}
 \subsubsection{Structure of the Hessian at the saddle point}
 \noindent
It follows from this that, at the saddle point $q_1=q^2$, for each replica $a$ the shifted Hessian can be written as:
\begin{equation}\label{eq:MatSplit}
 \frac{\tilde{\mathcal{M}}}{\sqrt{N}} = \frac{\mathcal{S}}{\sqrt{N}}+ \mathcal{D},
\end{equation}
where $\mathcal{S}$ is a stochastic matrix with the block structure:
\begin{equation}\label{eq:newMatrix} 
 \mathcal{S}= \begin{pmatrix}
               \mathcal{S}_0 & \mathcal{S}_{1/2}\\
               \mathcal{S}_{1/2}^T & \mathcal{S}_{1}
              \end{pmatrix},
\end{equation}
where the largest $(N-1-n) \times (N-1-n)$ block $\mathcal{S}_0$ is a GOE with $\sigma^2=p(p-1)$, $\mathcal{S}_{1/2}$ is an $(N-n-1) \times n$ block with iid Gaussian entries
\begin{equation}\label{eq:MatrixStochastic}
 \mathcal{S}_{1/2}=\left( 
\setlength{\arraycolsep}{.1pt}
        \begin{array}{ccc} n_{1M+1} &  \cdots  &n'_{1N-1} \\
n_{2M+1} &  \cdots  &n'_{2N-1} \\
  \cdots & \cdots & \cdots\\
  \cdots & \cdots & \cdots\\
   
  n_{M M+1} &  \cdots  &n'_{M N-1}\\
                 \end{array}
      \right),
\end{equation}
where the $n_{ij}$ have variance $\delta^2(q)=p(p-1)[1- c(q) q^{2p-4}]
$, while the $n'_{iN-1}$ have yet another variance:
\begin{equation}
 \Delta^2(q)=p(p-1)[1- c(q)].
\end{equation}
The conditioning thus reduces the fluctuations of these matrix elements with respect to the unconditioned case. The smaller $n \times n$ block  $\mathcal{S}_{0}$ has entries that are mutually correlated, with non-zero averaged contained in the 
deterministic matrix $ \mathcal{D}$. The latter has also a block structure $\mathcal{D}= \text{diag} \tonde{\mathcal{D}_0, \mathcal{D}_1}$, with $\mathcal{D}_0=0$ and 
\begin{equation}\label{eq:Deterministic}
\mathcal{D}_1=\left( 
\begin{array}{ccccc}
\setlength{\arraycolsep}{.1pt}
   \nu &  0  &  \cdots&  \cdots &0 \\
0 & \nu& \cdots&  \cdots  &0 \\
   0 &  0  & \nu&  \cdots &0 \\
  0 &  \cdots &  \cdots &\nu & 0\\
    0 & \cdots &  \cdots &0 & \mu\\
                     \end{array} 
      \right)
\end{equation}
with $\nu, \mu$ given in \eqref{eq:Nu}, \eqref{eq:Mu}. 
The conditional Hessian $\tilde{H} / \sqrt{N}$ is obtained after a shift with a diagonal matrix, see Eq. \eqref{eq:ShiftedHessian}. It is thus a shifted Gaussian matrix, perturbed with finite rank perturbation. Notice that in the annealed case (\emph{i.e.}, for $n \to 1$), only one special line and column remain (the last one). As we shall now see, from the point of view of the isolated eigenvalue these are indeed the only column and row that matter; thus, even at the level of the eigenvalue the quenched calculation reproduces the annealed one. 
 \subsubsection{Computation of the isolated eigenvalue}
 \noindent
In the large-$N$ limit, the bulk of the density of eigenvalues of $\tilde{\mathcal{M}}/\sqrt{N}$ is controlled by the GOE block, and is thus a centered semicircle. To discuss the stability of the stationary points, we need to compute the lower order corrections to this density of states, to determine whether there are isolated eigenvalues that become negative, inducing an instability. To this aim, we need to determine the poles of the resolvent of $\tilde{\mathcal{M}}/\sqrt{N}$ that lie on the real axis and are smaller than $ -2 \sqrt{p(p-1)}$. This requires to compute the trace of $(z- \tilde{\mathcal{M}}/\sqrt{N})^{-1}$. We focus on the contribution to the trace coming from the small $n \times n$ block of the resolvent; indeed, the corresponding matrix elements are the ones having non-zero overlap with the fixed minimum ${\bm \sigma}^0$, and thus only the poles of this part of the resolvent can be generated by the conditioning and can have eigenvectors with a non-zero component in the direction of the fixed minimum. The quantity to determine are therefore the poles of $ \langle \text{Tr} \grafe{1/N \cdot D(z)} \rangle$, where 
\begin{equation}\label{eq:Dz}
 D(z) \equiv z \hat{1}- \frac{{\mathcal{S}_1}}{\sqrt{N}}-\frac{1}{N} \mathcal{S}^T_{1/2} \tonde{z \hat{1}- \frac{{ \mathcal{S}_0}}{\sqrt{N}}}^{-1}  \mathcal{S}_{1/2},
\end{equation}
and where now the average is over the distribution of the entries of the matrix $\mathcal{S}$. Following the same step as in Ref. \cite{SpikedRepKacRice}, we find that the poles are solutions of the equation:
\begin{equation}\label{eq:Poles}
 \quadre{z- \nu - \delta^2 G_\sigma(z)}^{n-1} \tonde{z- \mu - \Delta^2 G_\sigma(z)}=0,
\end{equation}
where 
\begin{equation}\label{eq:ResolventGOE}
 G_\sigma(z)=\frac{z+ \sqrt{z^2-4 \sigma^2}}{2 \sigma^2}
\end{equation}
is the resolvent of a GOE matrix with variance $\sigma^2$. In particular, for $n \to 0$ we can focus on the solutions of $z- \mu - \Delta^2 G_\sigma(z)=0$, which satisfy:
\begin{equation}\label{eq:Eq1auto}
 z \tonde{1- \frac{1}{2}\frac{\Delta^2(q)}{\sigma^2}}-\mu(q, \epsilon, \epsilon_0)= \frac{1}{2}\frac{\Delta^2(q)}{\sigma^2}\sqrt{z^2- 4 \sigma^2}.
\end{equation}
We notice that for fixed $q$ and $\epsilon_0$, $\mu<0$ is a decreasing function of $\epsilon$: this already indicates that the additive part of the rank-1 perturbation is stronger for stationary points that are at higher energy, that are therefore more prone to an instability towards ${\bm \sigma}^0$. \\
Taking the square of \eqref{eq:Eq1auto}, we obtain a second order equation for $z$,
\begin{equation}
 z^2 \tonde{1- \frac{\Delta^2}{\sigma^2}}- 2 \mu \tonde{1- \frac{\Delta^2}{2 \sigma^2}}z+ \tonde{\mu^2+\frac{\Delta^4}{\sigma^2}}=0.
\end{equation}
Of the two solutions $z_{\pm}(q, \epsilon, \epsilon_0)$ of this equations (differing for the sign in front of the square root), only those that are real and satisfy
\begin{equation}\label{eq:ConditionIsolated}
 z_{\pm} \tonde{1- \frac{\Delta^2}{2\sigma^2}}-\mu \geq 0
\end{equation}
have to be retained, as they are consistent with the choice of the sign in front of the square root in \eqref{eq:ResolventGOE}, see Eq.~\eqref{eq:Eq1auto}. The point at which the equality holds in \eqref{eq:ConditionIsolated} correspond to the value of parameters for which the eigenvalue detaches from the lower edge of the support of the semicircle. For the values of the parameters $q, \epsilon$ and $\epsilon_0$ that we are interested in, we find that the relevant solution, whenever it exists, equals to $z_{+}$. Given this solution, the isolated eigenvalue is obtained from:
\begin{equation}\label{eq:IsolatedEigenvalue}
 \lambda_{0}(q, \epsilon, \epsilon_0)= z_{+}(q, \epsilon, \epsilon_0)- \sqrt{2} p \epsilon,
\end{equation}
which is equivalent to Eq. (4) in the main text. 

\subsection{Additional results on the complexity}\label{app:AdditionalResults}
\subsubsection{Quenched vs annealed average over the fixed minimum}
\noindent
The average over the disorder in Eq. (2) in the main text is conditioned to ${\bm \sigma}^0$, meaning: 
\begin{equation}\label{eq:Cond}
 \langle \cdot \rangle_0\equiv \Big\langle \cdot \hspace{.1cm}\Big| \grafe{
 \begin{subarray}{l}
 {\bf g}[{\bm \sigma}^0]=0,\\
  h[{\bm \sigma}^0]=\sqrt{2 N} \epsilon_0 \end{subarray}}
 \Big\rangle.
\end{equation}
The resulting complexity does not depend explicitly on ${\bm \sigma}^0$, and can therefore be trivially averaged with respect to the flat measure over stationary points of a given energy $\epsilon_0$, meaning that
\begin{equation}\label{eq:AnnealedSigma}
 \Sigma(\epsilon, q| \epsilon_0)= \lim_{N \to \infty} \frac{1}{N} \frac{1}{\mathcal{N}(\epsilon_0)} \int \mathcal{D} {\bm \sigma}^0 \langle \log \mathcal{N}_{{\bm \sigma}^0}(\epsilon, q| \epsilon_0) \rangle_0,
\end{equation}
where $\mathcal{D}{\bm \sigma}^0$ is the flat measure over stationary points with the right energy density,
\begin{equation}
 \mathcal{D}{\bm \sigma}^0 = d{\bm \sigma}^0  \delta(h[{\bm \sigma}^0]-\sqrt{2 N} \epsilon_0)\; \delta({\bf g}[{\bm \sigma}^0]),
\end{equation}
and $\mathcal{N}(\epsilon_0)$ is their total number. This corresponds to performing an \emph{annealed} average over the stationary point ${\bm \sigma}^0$, which is in fact equal to the \emph{quenched} average: 
\begin{equation}\label{eq:QuenchedSigma1}
 \lim_{N \to \infty} \frac{1}{N} \left\langle \frac{1}{\mathcal{N}(\epsilon_0)} \int \mathcal{D} {\bm \sigma}^0  \log \mathcal{N}_{{\bm \sigma}^0}(\epsilon, q| \epsilon_0)\right\rangle,
\end{equation}
where the average over the stationary points ${\bm \sigma}^0$ is performed \emph{prior} to the disorder average. The reason for the equivalence is that stationary points of the unconstrained $p$-spin landscape are typically orthogonal to each others, and thus  uncorrelated (this indeed also implies that the quenched and annealed complexity of the unconstrained $p$-spin coincide \cite{crisom92,crisom95,subag}). Indeed, the disorder average \eqref{eq:QuenchedSigma1} can be computed as: 
\begin{equation}
\begin{split}
 &\lim_{m \to 0} \left \langle [\mathcal{N}(\epsilon_0)]^{m-1} \int \mathcal{D} {\bm \sigma}^0  \log \mathcal{N}_{{\bm \sigma}^0}(\epsilon, q| \epsilon_0)\right \rangle =\\
 & \lim_{m \to 0} \left \langle \int \mathcal{D} {\bm \sigma}^{0} \prod_{k=1}^{m-1}\mathcal{D} {\bm \tau}^{k} \log \mathcal{N}_{{\bm \sigma}^0}(\epsilon, q| \epsilon_0)\right \rangle, 
 \end{split}
\end{equation}
and similarly to \eqref{eq:Full} this equals to 
\begin{equation}\label{eq:ReplicatedQuenchedZero}
\begin{split}
  \lim_{m \to 0}&\int d {\bm \sigma}^{0} \prod_{k=1}^{m-1}d {\bm \tau}^{k}  \mathcal{E}_{{\bm \sigma}^0,\vec{{\bm \tau}}}(\epsilon_0)\,p_{{\bm \sigma}^0, \vec{{\bm \tau}}}({\bf 0},  \epsilon_0) \cdot \\
  &\cdot \left \langle \log \mathcal{N}_{{\bm \sigma}^0}(\epsilon, q| \epsilon_0)  \Big| \grafe{
 \begin{subarray}{l}
 {\bf g}[{\bm \sigma}^0]=0,{\bf g}[{\bm \tau}^k]=0\\
  h[{\bm \sigma}^0]=h[{\bm \tau}^k]=\sqrt{2 N} \epsilon_0
  \end{subarray}} \right \rangle,
  \end{split}
\end{equation}
where $p_{{\bm \sigma}^0, \vec{{\bm \tau}}}({\bf 0},  \epsilon_0)$ is now the joint distribution of gradients and energy densities of ${\bm \sigma}^0$ and of the ${\bm \tau}^k$, and $\mathcal{E}_{{\bm \sigma}^0,\vec{{\bm \tau}}}(\epsilon_0)$ the expectation value of the corresponding determinant. This expression can be parametrized in terms of the overlaps $z_{kl}= {\bm \tau}^k \cdot {\bm \tau}^l$, $z_{k0}= {\bm \tau}^k \cdot {\bm \sigma}^0$ and $z_{ka}= {\bm \tau}^k \cdot {\bm \sigma}^a$, in addition to the previously introduced overlaps $q_{ab}$ and $q$, which is fixed. Performing first the saddle point over the overlaps between the ${\bm \tau}^k$ and ${\bm \sigma}^0$ one finds $z_{kl}=0=z_{k0}$ and $z_{ka}=0$; it follows that the disorder average in \eqref{eq:ReplicatedQuenchedZero} becomes independent of the ${\bm \tau}^k$ and reduces to 
\eqref{eq:Cond}, and thus \eqref{eq:ReplicatedQuenchedZero} reduces to \eqref{eq:AnnealedSigma}. 
\subsubsection{The complexity at $q_M(\epsilon_0)$}
\noindent
\begin{figure}[!htbp]
\includegraphics[width=.95\columnwidth]{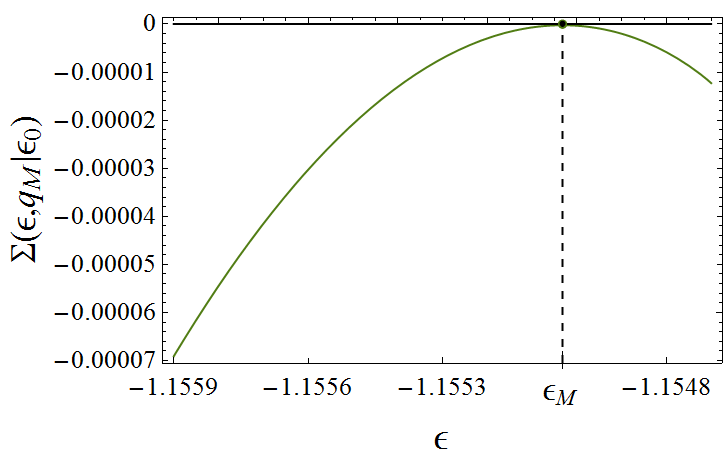}
    \caption{Complexity for $\epsilon_0=-1.158$ and $q=q_M(\epsilon_0)=0.83683$. The local maximum is at $\epsilon_M(\epsilon_0)=-1.1550< \epsilon_{\text{th}}(\epsilon_0)$, where $\Sigma=0$.}\label{fig:FigSupp1}
 \end{figure}%
The complexity is positive (implying that exponentially-many stationary points are present) only for $q_{\text{m}}(\epsilon_0)\leq  q \leq q_{\text{M}}(\epsilon_0)$, where $q_{\text{m}}(\epsilon_0)<0$. For each $q$ in this range, the stationary points with energy below the threshold are distributed over an extensive spectrum of $\epsilon$, which at $q_{\text{m}}$ and $q_{\text{M}}$ collapses to a single point. The lower boundary $q_{\text{m}}(\epsilon_0)$ varies very little with $\epsilon_0$, and $\overline{\epsilon}_{x=0}(q_{\text{m}}|\epsilon_0)=\epsilon_{\text{th}}$. At $q_{\text{M}}(\epsilon_0)$, instead, two different situations are possible: for the smaller values of $\epsilon_0$, the complexities $\Sigma(\epsilon, q|\epsilon_0)$ are increasing in the interval $\overline{\epsilon}_{x=0}(q|\epsilon_0) \leq \epsilon \leq \epsilon_{\text{th}}$ for any $q$, and $q_{\text{M}}(\epsilon_0)$ is the point at which $\overline{\epsilon}_{x=0}(q_{\text{M}}|\epsilon_0)=\epsilon_{\text{th}}$; for values of $\epsilon_0$ very close to the threshold, instead, the complexity is no longer monotonic but has a local maximum within the interval, and $q_{\text{M}}(\epsilon_0)$ is the latitude at which the local maximum touches zero, see Fig. \ref{fig:FigSupp1}. This implies that $\Sigma$ is negative everywhere (included above the threshold) except at one precise value of energy density, $\epsilon_{\text{M}}(\epsilon_0) \leq \epsilon_{\text{th}}$, where $\Sigma=0$. Note however that even for these $\epsilon_0$ one can find barriers at energies up to $\epsilon_{\text{th}}$, by focusing on small enough overlaps.

\subsubsection{The vanishing of the isolated eigenvalue.}
\noindent
Fig. 1 in the main text shows that minima appear first at the overlap $q^*(\epsilon_0)$, and in a small interval of $q \lesssim q^*(\epsilon_0)$ they coexist with saddles: the higher-energy points are saddles and the lower energy ones are minima, separated by a family of \emph{marginal} saddles (with one single zero mode), having finite complexity. At smaller values of $q$, all stationary points are minima. We now argue that, for any fixed $\epsilon_0$ (we henceforth drop the dependence on $\epsilon_0$), the iso-complexity curves $\overline{\epsilon}_x(q)$ satisfying 
\begin{equation}
 \Sigma(\overline{\epsilon}_x, q) \equiv x,
\end{equation}
for those values of $x$ for which they are non-monotonic below the threshold energy, have a local minimum at a point (say $q=q_x$) which is also the point at which the isolated eigenvalue vanishes, 
\begin{equation}
 \lambda_{0}(q_x, \overline{\epsilon}_x(q_x))=0.
\end{equation}
Indeed, at $q=q_x$ and $\epsilon=\overline{\epsilon}_x(q_x)$ it holds simultaneously:
\begin{equation}\label{eq:System1}
 \begin{cases}
  \frac{\partial}{\partial q} \Sigma(\epsilon; q)&=0\\
  \Sigma(\epsilon, q)& = x,
 \end{cases}
\end{equation}
where the first equation follows from $d \Sigma(\overline{\epsilon}_x(q), q)/ d q=0$, using that $d \overline{\epsilon}_x(q) / d q=0$. On the other hand, the isolated eigenvalue $\lambda_0$ vanishes whenever:
\begin{equation}\label{eq:System2}
 \mu(q, \epsilon)+ \Delta^2 G_\sigma(\lambda_0)- \sqrt{2} p \epsilon=0, 
  \end{equation}
where we used that $\lambda_0-\mu-\Delta^2 G_\sigma(\lambda_0)=0$. Eq.~\eqref{eq:System2} and the first of the Eqs.~\eqref{eq:System1} are both second order equations for $\epsilon$ at fixed $q$, and substituting the explicit expressions for the constants it can be shown that they are proportional to each others, and thus admit identical solutions. This fixes two curves $\overline{\epsilon}_{\pm}(q)$, one of which can be selected by imposing the consistency with the sign in front of the square root of the resolvent. Imposing the condition $\Sigma(\overline{\epsilon}_{\pm}(q), q)=x$, one selects the point $q_x$. 
\subsubsection{Comparison with the unconstrained complexity of index-1 saddles}
\begin{figure}[!htbp]
\captionsetup[subfigure]{labelformat=empty}
\subfloat[][\hspace{1.7 cm}(a)]{\includegraphics[width=\linewidth]{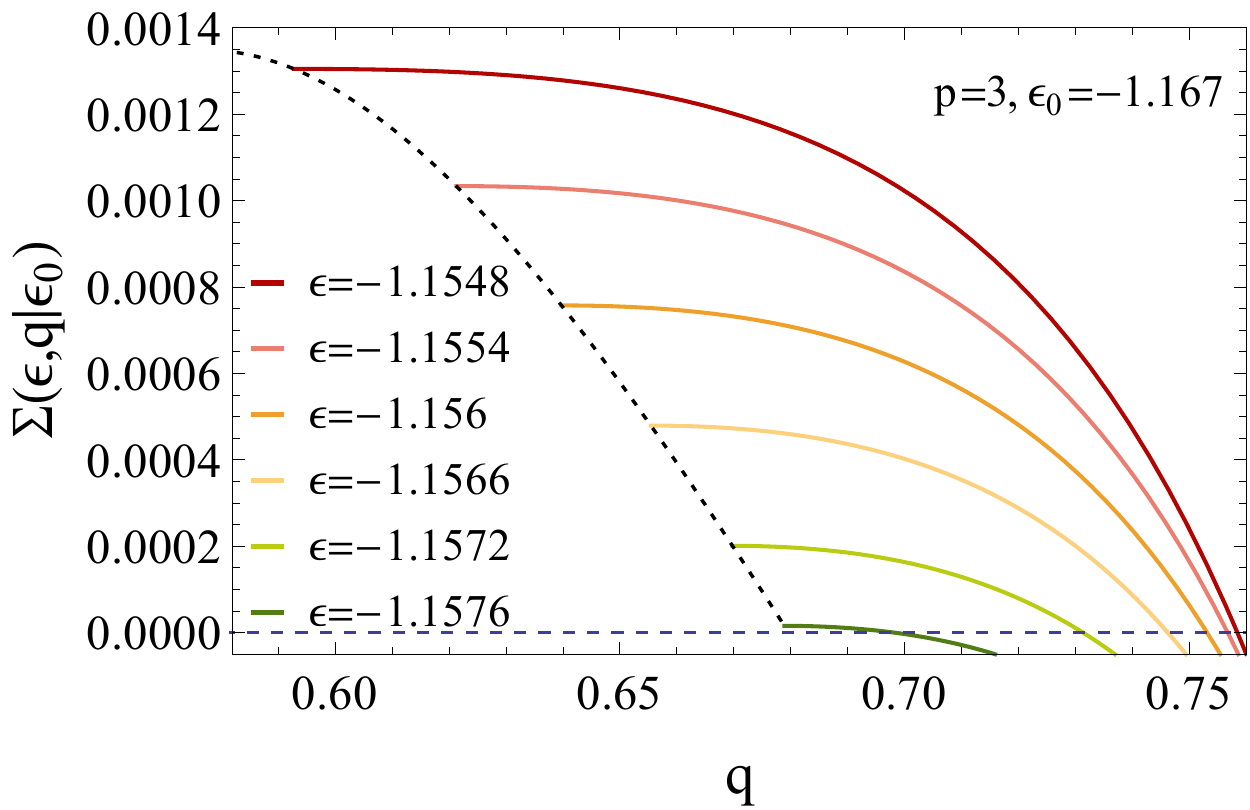}}\\
\subfloat[][\hspace{1.7 cm}(b)]{\includegraphics[width=\linewidth]{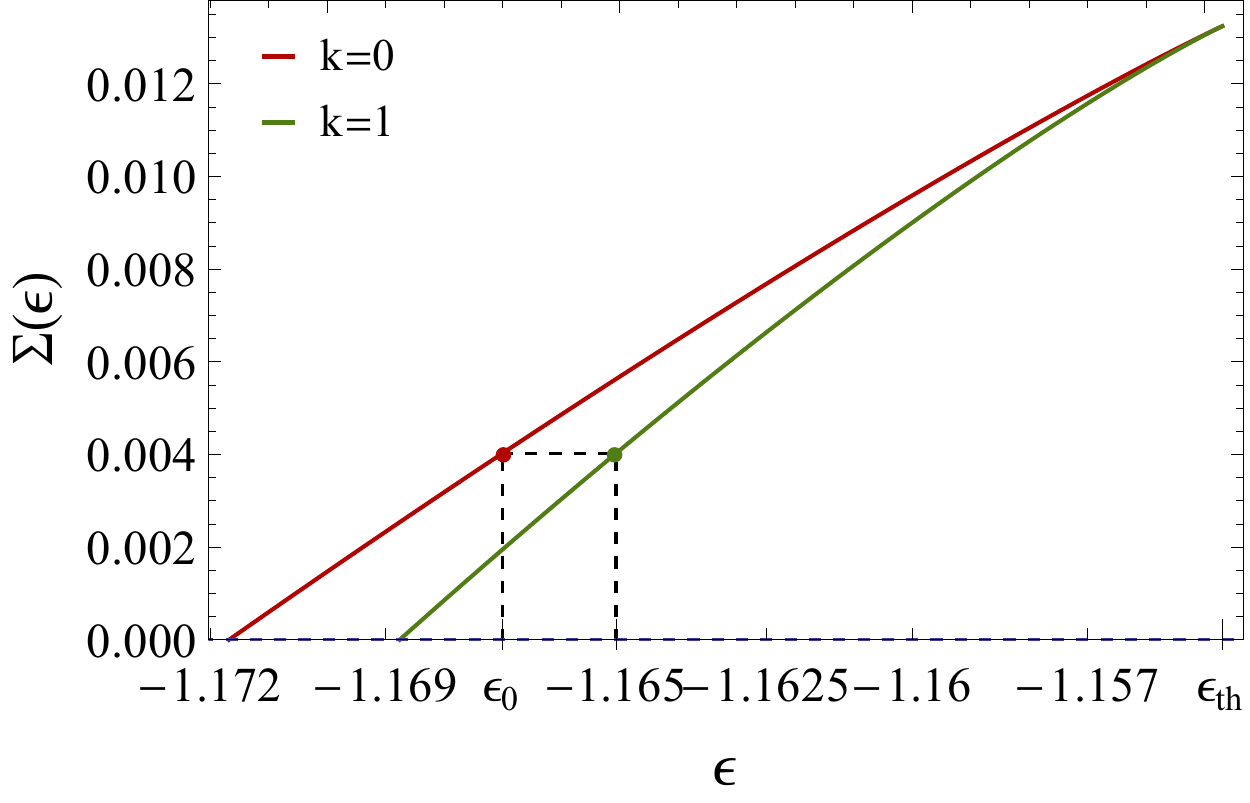}}
    \caption{(a) Complexity of the saddles at overlap $q$ from ${\bm \sigma}^0$. The curves reach their maximum at the overlaps corresponding to the marginal saddles with a single zero mode (they continue at smaller $q$ with a decreasing branch, not plotted, corresponding to minima).  The black dotted curve is the complexity of marginal saddles. (b) Unconstrained complexity of minima ($k=0$) and index-1 saddles ($k=1$). The red point identifies $\epsilon_0=-1.167$, the green point the energy of the index-1 saddles equally numerous with respect to the minima at energy  $\epsilon_0=-1.167$. }\label{fig:BenArousPlot}
 \end{figure}

\noindent
Fig. \ref{fig:BenArousPlot} shows a comparison between the complexity of the saddles at fixed overlap with ${\bm \sigma}^0$ and the total complexity of the minima and order-1 saddles of the $p$-spin landscape. It   shows that the saddles found with our calculation are not the index-1 saddles having the same complexity as the family of minima to which ${\bm \sigma}^0$ belongs (\emph{i.e.}, the minima with energy $\epsilon_0$). Rather, the saddles found in the vicinity of ${\bm \sigma}^0$ have higher energy. We remind that these are the properties of the \emph{typical} stationary points found at fixed $q$, \emph{i.e.}, of the most numerous ones, counted by $\Sigma(\epsilon, q|\epsilon_0)$. Rarer points with different stability properties should be present at the same latitudes: to determine their complexity, however, one has to perform large deviation calculations by conditioning explicitly on their index. We leave this computation for future work.

 \subsection{Zero-temperature Franz-Parisi potential}\label{app:FranzParisi}
 \noindent
 In this last section, we report the saddle point equations obtained when computing the free energy of a system constrained to be at fixed overlap $q$ from a given minimum of energy density $\epsilon_0\geq \epsilon_{\text{gs}}$. The computation is performed at zero temperature ($\beta' \to \infty$). The free energy is obtained as
\begin{equation}
F(q|\epsilon_0)=-\lim_{\beta' \to \infty}\lim_{n \to 0}\frac{1}{\beta'} \frac{S(q|\epsilon_0)}{n}
\end{equation}
where $S(q|\epsilon_0)$ is such that: 
\begin{equation}\label{eq:ActionRep}
 \begin{split}
 &\exp \quadre{NS(q|\epsilon_0)+o(N)}=\int d{\bf Q} \exp \quadre{NS({\bf Q})+ o(N)}=\\
 &\left \langle \int d{\bf s} \,
\text{exp}\tonde{-\beta' \sum_{\alpha=1}^n \sum_{i_1 < \cdots <i_p}J_{i_{1} \cdots i_p} s_{i_1}^\alpha \cdots s_{i_p}^\alpha}
\right \rangle_{0}.
 \end{split}
\end{equation}
The integral on the RHS is over configurations ${\bf s}$ on an hypersphere of radius $\sqrt{N}$ constrained 
to be at overlap $q$ from a given minimum at energy density $\epsilon_0$; the average is both on the random couplings, and on minima at energy $\epsilon_0$. In order to average over minima we introduce additional $m$ replicas and follow \cite{Monasson}: we select a given energy level by sampling from a Boltzmann measure at inverse temperature $\beta$, biased by the number of replicas $m$. The value of $m$ is not optimized over as it should be done to obtain the equilibrium energy, but it is chosen in such a way to select the energy density $\epsilon_0$ at will.
In particular, to select minima we work in the limit $\beta\rightarrow\infty$ and $m\rightarrow 0$ such that $\beta m$ is finite and chosen appropriately to select minima of energy $\epsilon_0$. The resulting integral on the LHS of \eqref{eq:ActionRep} is over $ (n + m)$ replicas and therefore over $ (n + m) \times (n + m) $ overlap matrices ${\bf Q}$, which parametrize the action as:
\begin{equation*}
 \begin{split}
  S({\bf Q})&=\frac{\beta^2}{4}\sum_{a,b=1}^m Q^p_{a,b}+\frac{\beta'^2}{4}\sum_{\alpha,\beta=1}^n Q^p_{\alpha,\beta}+\frac{\beta\beta'}{2}\sum_{a,\alpha}Q^p_{a,\alpha}\\
  &+\frac{1}{2}\log\det({\bf Q}).
 \end{split}
\end{equation*}
The overlap matrix ${\bf Q}$ has the following generic structure: it is replica-symmetric in the first $m \times m$ diagonal block, corresponding to the minimum of energy $\epsilon_0$, and has entries equal to $q_0$ everywhere except from the diagonal elements, that are set to $1$; it is either replica symmetric (or 1-step replica symmetry broken) in the second $n \times n$ diagonal block, which describes the system at fixed overlap $q$ with the minimum, with $1$ on the diagonal and $q'$ anywhere else (or $q'_1$ in the $\mu \times \mu$ diagonal sub-blocks and $q'_0$ anywhere else); finally, the elements belonging to the $n \times m$ off-diagonal rectangles are all set equal to $q$. Below, we determine the saddle point equations and compute the corresponding action in the two cases.

\subsubsection{The replica-symmetric case}
\noindent
%In the RS case the eigenvalues of the overlap matrix are $1-q_0$ with multiplicity $m-1$, $1-q'$ with multiplicity $n-1$, and the product of the last two eigenvalues is $[1+q_0(m-1)][1+q'(n-1)]-m n q^2$.
In the RS case, the action reads as follows:
\begin{equation}\label{eq:RSaction}
 \begin{split}
S_{RS}&=\frac{1}{4}[\beta^2(m+m(m-1)q_0^p)+\beta'^2(n+n(n-1)q'^p)+\\
&+2 m n\beta\beta'q^p]+\\
&+\frac{1}{2}\{(m-1)\log(1-q_0)+(n-1)\log(1-q')+\\
&+\log[(1+(m-1)q_0)(1+(n-1)q')-m n q^2]\}. 
 \end{split}
\end{equation}
The saddle point equation for $q_0$ gives
\begin{equation}
\frac{\beta m}{2}pq_0^{p-1}\hspace{-1 pt}=\frac{1}{\beta(1-q_0)}-\frac{\beta'(1-q'+n q')}{D}
\end{equation}
with 
\begin{equation}
 D=\beta' (1-q'+n q')\beta(1-q_0+m q_0)-m\beta \beta' n q^2,
\end{equation}
which for $n \to 0$ becomes:
\begin{equation}\label{eq:Saddleq0}
\beta(1-q_0)\beta(1-q_0+m q_0)=\frac{2}{p}.
\end{equation}
In the $\beta \to \infty$ limit, the product $\beta(1-q_0)$ remains finite and equals:
\begin{equation}
 \beta(1-q_0)= \frac{1}{2} \tonde{-\beta m+ \sqrt{(\beta m)^2 + \frac{8}{p}}},
\end{equation}
and the energy $\epsilon_0$ of the corresponding minimum can be written as:
\begin{equation}
 \epsilon_0=- \frac{1}{2} \tonde{(p-1) \beta(1-q_0)+ \frac{2}{p \beta(1-q_0)}}.
\end{equation}
These two equations fix $\beta m$ as a function of the chosen $\epsilon_0$.\\
The saddle point on $q'$ gives instead:
\begin{equation}
\frac{\beta' n}{2}pq'^{p-1}=\frac{1}{\beta'(1-q')}-\frac{\beta(1-q_0+m q_0)}{D}
\end{equation}
which in the $n\to 0$ limit reduces to
\begin{equation}
\frac{\beta'^2}{2}pq'^{p-1}=\frac{q'}{(1-q')^2}\left(1-\frac{q^2 m}{q'(1-q_0+m q_0)}\right) \ .
\end{equation}
In the limit $\beta'\rightarrow\infty$, we have $q'\rightarrow 1$ with $\beta'(1-q')$ finite, fixed by the equation:
\begin{equation}
\beta'^2(1-q')^2=\frac{2}{p}\left(1-\frac{q^2 \beta m}{\beta(1-q_0+m q_0)}\right) \ .
\end{equation}
This solution is stable for high values of the overlap $q$ and until 
\begin{equation}\label{eq:replicon}
\beta'^2 (1-q')^2=\frac{2}{p (p-1)} \ ,
\end{equation}
where the replicon eigenvalue of the hessian corresponding to this solution vanishes~\cite{crisom92}.

\subsubsection{The 1-step replica symmetry broken case}
\noindent
In this case the ${\bf Q}$ matrix has a 1RSB structure in the $n \times n$ block, with $q'_1$ on the diagonal $\mu \times \mu$ blocks and $q'_0$ in the rest of the matrix, except from the diagonal which is equal to $1$. 
%The corresponding eigenvalues are  $1-q_0$ with multiplicity $m-1$, $1-q'_1$ with multiplicity $n(1-1/\mu)$, $1-q'_1+\mu(q'_1-q'_0)$ with multiplicity $n/\mu-1$, and the product of the last two is $(1-q_0+m q_0)(1-q'_1+\mu(q'_1-q'_0)+n q'_0)-m n q^2$.
The action reads:
\begin{equation}\label{eq:RSBaction}
 \begin{split}
&S_{1RSB}=\\
&\frac{1}{4}[\beta^2(m+m(m-1)q_0^p)+\\
&+\beta'^2(n+n(\mu-1)q'^p_1+n(n-\mu)q'^p_0)+2m n \beta \beta' q^p]+\\
&+\frac{1}{2}\left[(m-1)\log(1-q_0)+n\tonde{1-\frac{1}{\mu}}\log(1-q'_1)+\right.\\
&\left.+\tonde{\frac{n}{\mu}-1}\log(1-q'_1+\mu(q'_1-q'_0))+\right.\\
&\left.+\log[(1-q_0+m q_0)(1-q'_1+\mu(q'_1-q'_0)+n q'_0)-m n q^2] \right].
\end{split}
\end{equation}
For $q=0$, this action reduces to the sum of one RS and one 1RSB actions with, respectively, inverse temperature $\beta$ and $m$ replicas, and inverse temperature $\beta'$ and a 1RSB structure with parameters $n$ and $\mu$: $S_{1RSB}(q=0)=S_{RS}(\beta,m)+S_{1RSB}(\beta',n,\mu)$. For arbitrary $q$, the saddle point on $q_0$ in the $n \to 0$ limit does not change with respect to the previous case, Eq. \eqref{eq:Saddleq0}. The saddle point equation for $q'_0$ and a combination of it with the equation for $q'_1$ give: 
\begin{equation}
\frac{p}{2}q'^{p-1}_0=\frac{1}{\beta'^2[1-q'_1+\mu(q'_1-q'_0)]^2}\left(q'_0-\frac{m}{1-q_0+m q_0} q^2\right)
\end{equation}
and
\begin{equation}
\frac{p}{2}(q'^{p-1}_1-q'^{p-1}_0)=\frac{q'_1-q'_0}{\beta'^2[1-q'_1+\mu(q'_1-q'_0)](1-q'_1)} \ .
\end{equation}
Finally, the saddle point on $\mu$ gives
\begin{equation}
 \begin{split}
  &0=\frac{q'^p_1-q'^p_0}{2}+\frac{1}{\beta'^2\mu^2}\log\left(\frac{1-q'_1}{1-q'_1+\mu(q'_1-q'_0)}\right)+\\
&+\frac{q'_1-q'_0}{\beta'[1-q'_1+\mu(q'_1-q'_0)]} \times  \\
&\times \left[\frac{1}{\mu\beta'}-\frac{1}{\beta'[1-q'_1+\mu(q'_1-q'_0)]}\left(q'_0-\frac{m q^2}{1-q_0+m q_0}\right)\right] \ .
 \end{split}
\end{equation}
For $\beta m$ and $q$ fixed, in the limit $\beta'\rightarrow\infty$ these equations have non-trivial solutions for $\beta'\mu$ finite (hence $\mu\rightarrow0$), $\beta'(1-q'_1)$ finite (hence $q'_1\rightarrow1$), and $q'_1-q'_0$ finite. For $q=0$, the usual 1RSB saddle point equations of the $p$-spin are recovered.\\
In both the RS and 1RSB case (Eq. \eqref{eq:RSaction} and \eqref{eq:RSBaction} respectively), for any fixed $\beta m$ the zero temperature FP potential, {\it i.e.} the minimal energy at fixed overlap from a typical minimum
at energy density $\epsilon_0$, is obtained by taking the derivative with respect to $n$ of the RS and 1RSB action as follows
\begin{equation}
\epsilon_{\text{FP}}(q|\epsilon_0)=-\lim_{\beta' \to \infty} \lim_{n \to 0}\frac{1}{\beta'}\frac{\partial S(q|\epsilon_0)}{\partial n},
\end{equation}
evaluating it in the $\beta \to \infty, m\to 0$ limit with $\beta m$ fixed. Using that $q'_1, q' \to 1$, we find in the RS case:
\begin{equation}\label{eq:FPrs}
\begin{split}
\epsilon_{\text{FP}}^{(RS)}=& -\frac{1}{2} \tonde{\beta m q^p + \frac{p}{2}\beta'(1-q')}\\
&-\frac{1}{2} \frac{\beta(1-q_0)+ \beta m(1-q^2)}{\beta' (1-q')(\beta(1-q_0)+ \beta m)},
\end{split}
\end{equation}
while in the 1RSB case we have:
\begin{equation}\label{eq:FPrsb}
\begin{split}
\epsilon_{\text{FP}}^{(1RSB)}=& -\frac{1}{2} \quadre{\beta m q^p + \frac{1}{2}\tonde{p \beta'(1-q'_1)+ \beta' \mu (1-q_0'^p)}}\\
&-\frac{1}{2} \frac{q_0'\tonde{\beta(1-q_0)+ \beta m}-\beta m q^2}{\tonde{\beta' (1-q'_1)+\beta' \mu (1-q_0')}(\beta(1-q_0)+ \beta m)}\\
&-\frac{1}{2 \beta' \mu} \log \tonde{\frac{\beta'(1-q'_1)+ \beta' \mu (1-q_0')}{\beta'(1-q_1')}},
\end{split}
\end{equation}
which reduces to the RS case when $q_0'=q_1' \to 1$. The expression \eqref{eq:FPrs} holds at high-enough $q$, until the condition \eqref{eq:replicon} is met, while \eqref{eq:FPrsb} holds at the smaller values of $q$. Substituting in these expression the solutions of the corresponding saddle point equations above, we obtain the result shown in Fig.4 of the main text. We remark that the FP has always a global minimum at $q=0$, which corresponds to the second configuration being at equilibrium at zero-temperature, independently from the fixed minimum of energy $\epsilon_0$; indeed, the corresponding energy \eqref{eq:FPrsb} equals to the ground-state energy of the unperturbed \emph{p}-spin. The local minimum of the zero-temperature FP potential is always attained at $q=1$: this corresponds to the second configuration being inside the state identified by the minimum of energy $\epsilon_0$, which at zero-temperature reduces to a single configuration (hence, the corresponding overlap equals to one). Finally, at the local maximum, where the potential coincides with $\overline{\epsilon}_{x=0}$, it holds $q'_0= q^2$, consistently with the saddle point solution for the overlap $q_1$ between the replicas in the Kac-Rice calculation (at that value of $q$, one finds indeed $q_0'=q_1$).      


\begin{thebibliography}{999}
\bibitem{fn:EnergyDensity} Stationary points of \eqref{eq:HampSpin} with intensive energy $\epsilon$ are also stationary points of the rescaled field with $h[{\bm \sigma}]= \sqrt{2 N} \epsilon$.
\bibitem{fn:optimal} Note that, as also discussed in the conclusion, the optimal barrier is the first one that is crossed by activated dynamics at low temperature. However, it is not necessarily the one that enables a full escape. This depends on the subsequent barriers to which it is connected to. If these are too high then it could be more convenient for the system to come back to the minimum and follow an alternative path.
\bibitem{fn:eMeth} Note that even in the cases when $\epsilon_{\text{M}}<\epsilon_{\text{th}}$ one can find barriers with energies up to $\epsilon_{\text{th}}$, by focusing on small
enough overlaps.
\bibitem{fn:top} The only information that one can gain from the FP potential is that its local maximum 
provides a lower bound to the height of the optimal barrier encountered by the system during thermal relaxation, \emph{i.e.}, 
when the system eventually decorrelates reaching zero overlap with the initial configuration \cite{montanarisemerjian,benarousj}.
\bibitem{fn:overlaps} In our case, in which the annealed approximation holds, the mutual overlap between optimal barriers is simply given by the square of the overlap $q$ between the minimum and the optimal barriers. For the three replica potential, where the annealed approximation does not hold, it is instead a more complicated function.
\bibitem{footnote1} In the following, we use latin indices to label the $n$ replicas, and greek indices to label the collection of the replicas and of the fixed stationary point at ${\bm \sigma}^0.$
\end{thebibliography}
\end{document}